\documentclass[iop,jphysa,preprint,article-title]{revtex4-2} % options: preprint or reprint, draft, linenumbers, ...
\usepackage[utf8]{inputenc} % use UTF-8 encoding in bibliography

\usepackage{cmap} % make ligatures etc. copy- and searchable
\usepackage[T1]{fontenc} % needed to have copy-able umlaute
\usepackage{lmodern}
\usepackage[scr]{rsfso} % calligraphic math font Formal Script

\usepackage{amsmath,amssymb,amsfonts}
\usepackage{mathtools,mleftright}
\usepackage{mathrsfs}
\usepackage{braket}
\usepackage{dsfont}% for unit matrix

\usepackage{graphicx}
\graphicspath{{./figures/}}
\newlength\figwidth
\usepackage{booktabs,tabularx,dcolumn}
\newcolumntype{d}[1]{D{.}{.}{#1}}

\usepackage{url,hyperref}
\usepackage[table,usenames,dvipsnames]{xcolor}
\hypersetup{colorlinks=true, linkcolor=BrickRed, urlcolor=blue!50!black, citecolor=blue!50!black}

\usepackage[capitalise]{cleveref}
\crefrangelabelformat{equation}{(#3#1#4)$-$(#5#2#6)}

\usepackage[normalem]{ulem}

\DeclareMathOperator\Imag{Im}
\DeclareMathOperator\Real{Re}
\DeclareMathOperator\Var{Var}
\DeclareMathOperator\tr{tr}

\renewcommand\vec[1]{{\boldsymbol #1}}

\newcommand\diff{\mathrm{d}}
\newcommand\e{\text{e}}
\renewcommand\i{\text{i}}
\renewcommand\geq\geqslant
\renewcommand\leq\leqslant
\renewcommand\ge\geqslant
\renewcommand\le\leqslant

\newcommand\kB{k_{\text{B}}}
\newcommand\ext{{}}% \text{ext}}
\newcommand\vD{v_\text{D}}

\begin{document}

\title{Memory effects in colloidal motion under confinement and driving}

\author{Arthur V. Straube}
\email{straube@zib.de}
\affiliation{
  Zuse Institute Berlin,
  Takustr. 7,
  14195 Berlin, Germany
}
\affiliation{
  Freie Universität Berlin,
  Fachbereich Mathematik und Informatik,
  Arnimallee 6,
  14195 Berlin, Germany
}

\author{Felix Höf{}ling}
\email{f.hoefling@fu-berlin.de}
\affiliation{
  Freie Universität Berlin,
  Fachbereich Mathematik und Informatik,
  Arnimallee 6,
  14195 Berlin, Germany
}
\affiliation{
  Zuse Institute Berlin,
  Takustr. 7,
  14195 Berlin, Germany
}

\date{\today}

\begin{abstract}
  The transport of individual particles in inhomogeneous environments is complex and exhibits non-Markovian responses.
  The latter may be quantified by a memory function within the framework of the linear generalised Langevin equation (GLE).
  Here, we exemplify the implications of steady driving on the memory function of a colloidal model system for Brownian motion in a corrugated potential landscape, specifically, for one-dimensional motion in a sinusoidal potential.
  To this end, we consider the overdamped limit of the GLE, which is facilitated by separating the memory function into a singular (Markovian) and a regular (non-Markovian) part.
  Relying on exact solutions for the investigated model, we show that the random force entering the GLE must display a bias far from equilibrium, which corroborates a recent general prediction.
  Based on data for the mean-square displacement obtained from Brownian dynamics simulations, we estimate the memory function for different driving strengths and show that already moderate driving accelerates the decay of the memory function by several orders of magnitude in time.
  We find that the memory may persist on much longer timescales than expected from the convergence of the mean-square displacement to its long-time asymptote.
  Furthermore, the functional form of the memory function changes from a monotonic decay to a non-monotonic, damped oscillatory behaviour, which can be understood from a competition of confined motion and depinning.
  Our analysis of the simulation data further reveals a pronounced non-Gaussianity, which questions the Gaussian approximation of the random force entering the GLE.
\end{abstract}

\maketitle

\section{Introduction}

The transport of individual particles in inhomogeneous environments is complex and exhibits non-Markovian responses, i.e., memory effects, which reflect the influence of the past trajectory on the future motion of the particle.
Typically, these memory effects depend on the timescales that are resolved in an experiment and they disappear at timescales longer than the slowest relaxation of the system.
One goal of data-driven modelling is to develop effective stochastic models for the particle motion which encapsulate the interaction with the environment in generalised transport coefficients rather than explicitly including the details of the environment.

Mathematically, the non-Markovianity manifests itself in a non-trivial functional form of the corresponding transport propagator, where the deviations from the standard (Markovian) form appear as a generalisation of the diffusion coefficient $D(k,\omega)$ that depends on a wavenumber $k$ and a frequency $\omega$ \cite{Hansen:SimpleLiquids,BoonYip:Molecular,Hoefling:RPP2013}.
Here, we consider the long-range transport of a single particle, which is described by the small-wavenumber regime ($k \to 0$)
and which is encoded in the velocity autocorrelation function (VACF) $Z(t)$ or, equivalently, the mean-square displacement (MSD);
in particular, $Z(t)$ and $D(k \to 0,\omega)$ are related to each other via a one-sided Fourier transform.
Deviations from a linear increase of the MSD (as function of the lag time $t$) indicate non-Markovian behaviour,
which typically approaches diffusive behaviour, i.e., a linear increase, at long times.

A widely employed framework for such non-Markovian behaviour of the MSD is the generalised Langevin equation (GLE, cf.\ \cref{eq:GLE}), at the heart of which is a generalised friction $\zeta(t)$, termed memory function or memory kernel.
Given $\zeta(t)$, the GLE generates particle trajectories with certain statistics and links, \emph{inter alia,} $\zeta(t)$ and the MSD.
The GLE has far-reaching applications, for example, in microrheology experiments, where one follows the motion of a probe particle in a complex medium to determine the MSD and to conclude, via the corresponding frequency-dependent friction, on the visco-elastic, mechanical responses of the medium \cite{Puertas:JPCM2014,Waigh:RPP2016,Rigato:NP2017}.

This phenomenological, top-down view is based on the observation of correlation functions or related quantities (e.g., the VACF or the MSD), whereas a complementary, bottom-up approach to the GLE is rooted in statistical mechanics and employs the projection operator techniques due to Zwanzig, Mori and others \cite{Zwanzig:1960, Mori:1965, Kubo:RPP1966}, see Ref.~\cite{Schilling:PR2022}  for a review.
A typical application is the projection on a suspended particle and thereby coarse-graining its interactions with a complex solvent \cite{Shea:SM2024,Milster:JCP2024};
recent progress was made on the non-linear GLE \cite{Glatzel:EPL2021,Ayaz:PRE2022,Vroylandt:EPL2022,DiCairano:JPC2022}
and on GLEs under non-equilibrium conditions \cite{Meyer:JCP2017,Netz:JCP2018,Jung:SM2021,Doerries:JSM2021,Koch:PRR2024,Jung:2023}.
In essence, the GLE may be seen as a means to encapsulate a large number of unresolved degrees of freedom in the memory function.
This apparent reduction of the complexity comes in general at the price of introducing non-Markovian dynamics, which, however, may be amenable to approximations depending on the specific physical system at hand.
If memory effects are entirely negligible, the Markovian approximation of the GLE reduces to the well-known Langevin equation, e.g., for the free diffusion of a Brownian particle on sufficiently long timescales.

It is a formidable challenge to parameterise the GLE from observations of a given complex system,
that is, estimating $\zeta(t)$ from data for the MSD or related correlation functions.
Today, several techniques are available for this task, which may roughly be divided into methods
based on few-parameter families of ansatz functions to represent the memory function
\cite{Gottwald:JCP2015, Daldrop:PRX2017, McKinley:SJMA2018},
and into ansatz-free methods in the time domain
\cite{Jung:JCTC2017, Meyer:EPL2020, Kowalik:PRE2019, Straube:CP2020}
or, via Fourier transforms, in the frequency domain \cite{Tassieri:NJP2012,RivasBarbosa:PCCP2020,Nishi:SM2018,Straube:CP2020}.
Notable recent advances follow a different route and rely on memory estimation using bare trajectory data as input instead of correlation functions \cite{Vroylandt:PNAS2022, Lapolla:PRR2021}.

Here, we consider the non-equilibrium dynamics of a Brownian particle in a corrugated potential landscape under external driving.
Experimentally, colloidal model systems provide a unique control over both the driving force and the environment,
where the latter may be realised as a potential landscape \cite{Evstigneev-etal:PRE2008,Straube-Tierno:EPL2013,Straube:SM2014,Juniper-etal:PRE2016,Juniper:NJP2017,Stoop-etal:NL2018,Stoop-etal:PRL2020}, as a structured substrate \cite{Ma:SM2015,Su:JCP2017,Choudhury:NJP2017} or a complex liquid \cite{Berner:NC2018}.
Furthermore, a considerable body of theoretical work is available for such systems (e.g., \cite{Reimann:PRE2002,Evstigneev-etal:PRE2008,Lapolla:FP2019}),
including also driven transport in random environments \cite{LeitmannJPAMT2018,LeitmannPRL2017,LeitmannPRL2013}.

Specifically, we study the overdamped motion of a Brownian particle driven by a constant external force $f$ across a corrugated potential landscape, which is of sinusoidal form with wave length $\lambda$ and wave number $k = 2\pi/\lambda$.
The corresponding overdamped Langevin equation for the particle's position $x(t)$ reads
\begin{align}
  \zeta_0 \dot{x}(t) = f - U'(x)  + \eta(t), \qquad U(x) = U_0(1- \cos kx)\,,
  \label{eq:Lang-BP}
\end{align}
where $\zeta_0$ denotes particle's friction constant in the absence of the potential
and we have assumed that the force points perpendicular to the grooves of the landscape and the $x$-axis is aligned with the force, rendering the problem essentially one-dimensional.
The random force $\eta(t)$ is a Gaussian white noise with mean $\langle\eta(t)\rangle = 0$
and covariance $\langle \eta(t) \eta(t') \rangle = 2 \kB T \zeta_0 \delta(t-t')$ in agreement with the fluctuation--dissipation relation for the solvent temperature $T$.

In equilibrium ($f=0$), the confining potential slows down Brownian motion at long times compared to free diffusion \cite{Lifson:JCP1962, Festa:PA1978}.
In terms of the MSD, there is a crossover between
diffusive regimes at short and long times with diffusion constants $D_0 = \kB T /\zeta_0$ and $D_\infty < D_0$, respectively.
Out of equilibrium, $f>0$, the presence of the driving force breaks the spatial inflection symmetry at $x=0$ and the system exhibits a non-vanishing drift velocity
$\vD = \lim_{t \to \infty} \langle \Delta x(t)\rangle/t$ in terms of the displacement $\Delta x(t):=x(t)-x(0)$.
Fluctuations of the position around this mean drift give rise to dispersion and
one defines the MSD as the variance $\delta x^2(t) := \langle \Delta x(t)^2 \rangle - \langle \Delta x(t) \rangle ^2$
with a corresponding long-time diffusivity $D_{\infty}=\lim_{t \to \infty} \delta x^2(t) /(2t)$.
Exact expressions for $\vD(f)$ and $D_\infty(f)$ are available in the literature \cite{Stratonovich:1967, Reimann:PRL2001, Reimann:PRE2002} and are quoted below [cf.\ \cref{eq:vD,eq:D-inf}].

Here, we will show that the dynamics due to \cref{eq:Lang-BP} can be reproduced by means of an overdamped version of the GLE equation, which reads [cf.\ \cref{eq:GLE-overd}]:
\begin{align}
    \zeta_0 \dot{x}(t) = f - \int_0^t \zeta_\text{reg}(t-s) \, \dot{x}(s) \, \diff s +  \xi(t)\,;
    \label{eq:GLE-overd-1d}
\end{align}
such that the confinement due to the potential $U(x)$ is encapsulated in the regular, continuous part $\zeta_\text{reg}(t)$ of the memory function.
The latter depends on the potential, but also on the driving force $f$, which is a well-known fact \cite{Daldrop:PRX2017, Post:JCTC2022}.
There are recent avenues \cite{Glatzel:EPL2021,Ayaz:PRE2022,Vroylandt:EPL2022} to develop non-equilibrium GLEs which employ the equilibrium memory function (as given for $f=0$), which comes at the price that the velocity enters the convolution integral non-linearly and the particle is subject to an effective systematic force, given by the potential of mean force and depending on the both $U(x)$ and $f$. Here, we will stick to the linear GLE.

\section{Non-equilibrium GLE and its memory function}
\label{sec:noneq-GLE}

For a colloidal bead of mass $m$ at position $\vec r(t)$ moving in $d$-dimensional space with velocity $\vec v(t)=\dot{\vec r}(t)$
and experiencing an external, constant force $\vec f_\ext$, the linear (in $\vec v$) GLE may be written as \cite{Kubo:RPP1966, Kubo:Book1991, Schilling:PR2022}
\begin{align}    
  m \dot{\vec v}(t) =  \vec f_\ext - \int_0^t \zeta(t-s; \vec f) \, \vec v(s) \,\diff s +
  \vec\xi_\vec f(t) \,.
  \label{eq:GLE}
\end{align}
Due to the force $\vec f_\ext$, the isotropic symmetry is broken and non-vanishing mean values of vectorial observables are possible.
In addition, the system is driven out of equilibrium in such a way that we can still expect the existence of a stationary statistical ensemble.
In particular, the latter permits a non-zero drift velocity $\vec \vD = \langle \vec v(t) \rangle$ and it defines the temperature $T$ via the variance $\Var(v_i) = \langle (\delta v_i)^2 \rangle = \kB T/m$ for every Cartesian component $\delta v_i$ of $\delta \vec v(t) := \vec v(t) - \vec \vD$.

The key ingredients of \cref{eq:GLE} are the memory function $\zeta(t; \vec f)$ and the random force $\vec{\xi}_\vec f(t)$, both of which depend on $\vec f$.
The random force may be biased for strong driving, i.e., in the non-linear response regime, and one expects a non-vanishing mean \cite{Shea:SM2024,Koch:PRR2024}. Therefore, we define
\begin{equation}
  \overline{\vec \xi}_{\vec f} := \langle \vec{\xi}_{\vec f}(t) \rangle \,.
  \label{eq:mean-noise}
\end{equation}%
Furthermore, $\vec \xi_{\vec f}(t)$ is approximated as a coloured Gaussian noise such that its covariance satisfies the fluctuation--dissipation relation,
\begin{equation}
  \langle \delta \vec{\xi}_\vec f(t) \otimes \delta \vec{\xi}_\vec f(t')\rangle
    = \kB T \zeta(|t-t'|; \vec f)
  \label{eq:coloured-noise}
\end{equation}
with $\delta \vec \xi_{\vec f}(t) = \vec \xi_{\vec f}(t) - \overline{\vec\xi}_{\vec f}$;
in the absence of isotropy, $\zeta(t;\vec f)$ is $d\times d$ matrix-valued.

The stationary drift velocity may be obtained from the GLE as the mean response of the bead to the driving force.
Taking the ensemble average of \cref{eq:GLE} in the long-time limit yields
\begin{equation}
  0 =  \vec f_\ext - \lim_{t\to \infty} \int_0^t \zeta(t-s; \vec f) \, \vec \vD(\vec f) \, \diff s
  + \overline{\vec\xi}_{\vec f}
\end{equation}
and thus
\begin{equation}
  \vec \vD(\vec f) = \zeta_\infty(\vec f)^{-1} (\vec f_\ext + \overline{\vec\xi}_{\vec f}) \,,
  \label{eq:drift-noise}
\end{equation}
in terms of the macroscopic, static friction coefficient
$\zeta_\infty(\vec f) := \int_0^\infty \zeta(s;\vec f) \, \diff s$.
In the linear response regime, for small driving force, it holds $\vD(\vec f) \propto \vec f$, which implies that $\zeta(t)$ becomes independent of $\vec f$ in this regime and that the noise is unbiased, $\overline{\vec\xi}_{\vec f \to 0} = 0$.

The long-range transport properties beyond $\vec \vD$ are encoded in the (matrix-valued) velocity autocorrelation function (VACF),
$Z(t; \vec f) = \langle \delta \vec v(t) \otimes \delta \vec v(0) \rangle$,
which is naturally expected to depend on the driving force.
Defining $Z(t; \vec f)$ in terms of the velocity fluctuation $\delta \vec v(t)$ instead of the velocity $\vec v(t)$ itself is the suitable generalisation to the non-equilibrium situation.
Building the corresponding scalar product for \cref{eq:GLE} and averaging yields the evolution equation of the VACF:
\begin{align}
  m \dot Z(t;\vec f)
    &=  \vec f_\ext \cdot \langle \delta \vec v(0) \rangle
      - \int_0^t \zeta(t-s;\vec f) \, \langle [\delta \vec v(s) + \vec \vD(\vec f)] \cdot \delta \vec v(0) \rangle \, \diff s
      + \langle \vec\xi_\vec f(t) \cdot \delta \vec v(0) \rangle  \notag \\
  &= - \int_0^t \zeta(t-s;\vec f) \, Z(s;\vec f) \, \diff s \,,
  \label{eq:VACF}
\end{align}
where we have exploited stationarity, $\langle \delta \vec v(t) \rangle = 0$ for any $t \geq 0$, and the independence of $\vec \xi_\vec f(t)$ and $\delta \vec v(0)$.
We emphasise that \cref{eq:VACF} has the same form as in equilibrium, which is due to the use of $\delta \vec v$ for the definition of $Z(t; \vec f)$.
In the following, we will often not indicate the explicit force dependence of the occurring quantities to avoid a cluttering of the notation.

\Cref{eq:VACF} is a linear integro-differential equation with the initial value $Z(0)= \kB T / m$.
It reduces to an algebraic equation for the one-sided Fourier (or: Fourier--Laplace) transform
$\hat Z(\omega) := \int_0^\infty \! \e^{\i \omega t} Z(t) \, \diff t$
with complex frequency $\omega$ such that $\Imag \omega > 0$.
This equation is straightforward to solve for $\hat Z(\omega)$, yielding
\begin{align}
  \hat{Z}(\omega)=\kB T [-\i\omega m+\hat{\zeta}(\omega)]^{-1} \,.
  \label{def:memory}
\end{align}

Relation \eqref{def:memory} provides a one-to-one link between the (complex matrix-valued) memory kernel $\hat\zeta(\omega)$ and the VACF.
Given $Z(t)$ and thus $\hat Z(\omega)$, \cref{def:memory} may also be considered as the defining relation for $\hat\zeta(\omega)$,
which uniquely implies the memory function $\zeta(t)$ as its (real matrix-valued) Fourier backtransform,
\begin{equation}
  \zeta(t) = \frac{2}{\pi} \int_0^\infty \cos(\omega t) \, \Real \hat \zeta(\omega) \, \diff \omega \,.
\end{equation}
In particular, $\hat\zeta(\omega)$ may be used to postulate \cref{eq:GLE} as a stochastic process that generates the prescribed $Z(t)$.
The fact that $Z(t)$ is a correlation function also guarantees the existence of a coloured noise $\vec \xi(t)$
with correlator $\kB T \zeta(t)$, so that the fluctuation--dissipation relation \eqref{eq:coloured-noise} is satisfied \cite{Teschl:MathMeth, Franosch:JPA2014}. % FIXME add Franosch & Voigtmann
This noise can be shifted suitably to satisfy \cref{eq:mean-noise}, which does not change the covariance in \cref{eq:coloured-noise}.

The standard Langevin equation,
\begin{equation}
  m \dot{\vec v}(t) + \zeta_0 \vec v(t) = \vec f + \vec\xi(t) \,,
  \label{eq:langevin}
\end{equation}
is contained in \cref{eq:GLE} as the special case where $\zeta(t)=\zeta_0 \mathds{1} \delta_+(t)$
is given by the identity matrix $\mathds{1}$ and a half-sided Dirac $\delta$-function $\delta_+(t)$ on $t \ge 0$;
the latter is understood in the sense that $\int_0^{\infty} g(t) \, \delta_+(t) \,\diff t = g(0)$ for continuous test functions $g(t)$.
Correspondingly, the noise $\vec\xi(t)$ is Gaussian white noise, independent of $\vec f$, such that
$\langle \vec\xi(t) \rangle =0$ and
$\langle \vec{\xi} (t) \otimes \vec{\xi}(t')\rangle = 2 \kB T \zeta_0 \mathds{1} \delta(t-t')$
with $\delta(t) = \delta_+(|t|) / 2$.
Solving \cref{eq:VACF} for this case yields $Z(t) = Z(0)\exp{(-\gamma_0|t|)}$ with the relaxation rate $\gamma_0 = \zeta_0/m$.
In the frequency domain, this corresponds to
\begin{equation}
  \hat{Z}(\omega) = \frac{\kB T}{-\i\omega m+\zeta_0} \,\mathds{1} \,,
\end{equation}
and, by comparison to \cref{def:memory}, $\hat{\zeta}(\omega)=\zeta_0 \mathds{1}$ reduces to a constant, Markovian friction.
One concludes that any contribution to $\zeta(t)$ beyond this singular part
(equivalently, any non-exponential decay of the VACF, $Z(t)$) indicates the presence of memory effects.

\section{Overdamped limit of the GLE}
\label{sec:overdamped}

\subsection{Derivation}

For application to overdamped dynamics, some modifications of the GLE and the definition of the memory function are needed.
We proceed in two steps: First, we explicitly split off the singular, Markovian part of the memory kernel,
which we assume to be also isotropic:
\begin{align}
    \zeta(t) = \zeta_0 \mathds{1} \delta_+(t) + \zeta_\text{reg}(t)\,, \label{eq:zeta-repres}
\end{align}
such that $\zeta_\text{reg}(t)$ is continuous as $t \to 0$
(see Ref.~\citenum{Bauer:EPJST2010} for a similar splitting of the VACF).
This turns \cref{eq:GLE} into
\begin{equation}
  m \dot{\vec v}(t) + \zeta_0 {\vec v}(t)= \vec f - \int_0^t \zeta_\text{reg}(t-s) \, \vec v(s) \diff s + \vec\xi_\vec f(t) \,.
\end{equation}
In particular, one explicitly sees that the standard Langevin equation, \cref{eq:langevin} is memoryless, $\zeta_\text{reg}(t)=0$, or Markovian.
Second, we move to the overdamped regime by the usual assumption that the inertial force is negligible compared with the friction force, $m \dot {\vec v}(t) \ll \zeta_0 \vec v(t)$. This implies that the inertial relaxation time $\gamma_0^{-1}$ is small compared to the timescales of interest, $t \gg \gamma_0^{-1}$.
Equivalently, in the frequency domain, we require that
\begin{align}
    |\omega| \ll \gamma_0=\frac{\zeta_0}{m}\,.
    \label{eq:omega-large}
\end{align}
The notion of the overdamped limit is such that the condition $t \gg \gamma_0^{-1}$ is fulfilled for all $t > 0$, which is realised by taking the limit $\gamma_0 \to \infty$, or, equivalently, by letting $m \to 0$ at constant $\zeta_0$.
The overdamped limit of \cref{eq:GLE} follows to be
\begin{align}    
    \zeta_0 \vec v(t) = \vec f - \int_0^t \zeta_\text{reg}(t-s) \, \vec v(s) \, \diff s + \vec\xi_\vec f(t) \,;
  \label{eq:GLE-overd}
\end{align}
the equilibrium form ($\vec f=0$) of this overdamped GLE was derived previously for the monomer dynamics in a Rouse chain \cite{Panja:JSM2010,Panja:JSM2010a} and was employed in the context of anomalous diffusion \cite{Miyaguchi:PRR2022}.
Time integration yields a closed stochastic integral equation for the displacement $\Delta\vec r(t) = \int_0^t \vec v(s)\, \diff s$:
\begin{align}
  \Delta \vec r(t) = \zeta_0^{-1} (\vec f +\overline{\vec\xi}_{\vec f})t
    - \zeta_0^{-1} \int_0^t \zeta_\text{reg}(t-s) \, \Delta \vec r(s) \, \diff s + \vec R(t) \,,
  \label{eq:GLE-displ}
\end{align}
which is driven by the centred Gaussian process
$\vec R(t) := \zeta_0^{-1} \int_0^t \, [\vec \xi(s) - \overline{\vec \xi}_{\vec f}] \, \diff s$.
This process has vanishing mean, $\langle \vec R(t) \rangle = \vec R(0) = 0$, and, recalling \cref{eq:coloured-noise}, is characterised by the covariance
\begin{align}
  \langle \vec R(t) \otimes \vec R(t')\rangle &=
  \zeta_0^{-2} \int_0^t \int_0^{t'} \kB T \bigl[2 \zeta_0 \mathds{1} \delta(s-s') + \zeta_\text{reg}(|s-s'|) \bigr]
  \diff s \,\diff s' \notag \\
%   &= 2 D_0 \mathds{1} \min(t,t') + D_0 \zeta_0^{-1}\int_0^{\min(t,t')} [\min(t,t')-s] \, \zeta_\text{reg}(s) \, \diff s \notag \\
%   & \qquad + D_0 \zeta_0^{-1} \int_0^{\max(t,t')} \min(\max(t',t)-s,t,t') \, \zeta_\text{reg}(s) \, \diff s
  &= 2 D_0 \mathds{1} \min(t,t') + D_0 C_\text{reg}(t) + D_0 C_\text{reg}(t') - D_0 C_\text{reg}(|t-t'|)\,,
  \label{eq:R-noise}
\end{align}
given in terms of $D_0 = \kB T/ \zeta_0$ and the $\vec f$-dependent correlation function
\begin{equation}
  C_\text{reg}(t) := \zeta_0^{-1} \int_0^t (t-s) \,\zeta_\text{reg}(s) \,\diff s \,,
  \label{eq:Creg}
\end{equation}
which is analogous to a familiar relation for the VACF [see below, \cref{eq:msd-via-Z(t)}].
The contribution $2 D_0 \mathds{1} \min(t,t')$ refers to free Brownian motion, i.e., a scaled Wiener process.
(For the derivation of \cref{eq:R-noise}, it is helpful to perform the integrals over $s$ and $s'$ first for $t' \geq t$; then, the term $D_0 C_\text{reg}(t)$ corresponds to the case $s > s'$ and the last two terms result from $s < s'$.)

\subsection{Velocity autocorrelation function}

The relation between the memory function and the VACF is given by the overdamped limit of \cref{eq:VACF},
but may also be obtained from \cref{eq:GLE-overd} directly:
\begin{align}    
  \zeta_0 Z(t) = - \int_0^t \zeta_\text{reg}(t-s) \, Z(s) \, \diff s  \,.
  \label{eq:VACF-overd}
\end{align}
This is a linear Volterra equation of the second kind in the function $Z(t)$; in particular, \cref{eq:VACF-overd} does not contain the time derivative $\dot Z(t)$, which is present in \cref{eq:VACF}.
Performing the Fourier-Laplace transform of \cref{eq:VACF-overd} or, alternatively putting
$\hat{\zeta}(\omega) = \zeta_0 \mathds{1} + \hat{\zeta}_\text{reg}(\omega)$ in \cref{def:memory} and letting $m \to 0$, one obtains:
\begin{align}
  \hat{Z}(\omega)= D_0 [\mathds{1}+\hat{\zeta}_\text{reg}(\omega)/\zeta_0]^{-1} \,.
  \label{def:memory-overd}
\end{align}
which may serve as the overdamped version of the definition of the memory kernel, analogously to \cref{def:memory}.

The reference case of unconfined Brownian motion of an overdamped particle, $\zeta_0 \dot{\vec r}(t) = \vec f + \vec\xi(t)$, corresponds to the memoryless situation, $\zeta_\text{reg}(t)=0$ [\cref{eq:GLE-overd}].
Then, the noise $\vec R(t)$ in \cref{eq:GLE-displ} simplifies to a Wiener process.
For the VACF, it follows from \cref{def:memory-overd} that $\hat{Z}(\omega)=D_0 \mathds{1}$ and, in the time domain, we have $Z(t) = D_0 \mathds{1} \delta_+(|t|)$.
This result coincides with the overdamped limit for the underdamped particle considered above, $Z(t) = \lim_{\gamma_0 \to \infty} D_0 \gamma_0 \mathds{1} \exp(-\gamma_0|t|)$.

\subsection{Mean-square displacement}

Given data for $Z(t)$, \cref{eq:VACF-overd} may be used to estimate the regular part of the memory function, $\zeta_\text{reg}(t)$.
For overdamped dynamics, however, the displacement $\Delta \vec r(t)$ is naturally observed in both experiments and simulations.
Thus, we wish to connect the memory function $\zeta_\text{reg}(t)$ of the overdamped GLE and the MSD, the latter being defined as $\delta r^2(t) = \langle |\Delta \vec r(t)|^2\rangle$.
Here and in the following, we will either consider isotropic motion in $d$-dimensional space or the motion along a fixed Cartesian axis ($d=1$); in particular, the matrix-valued VACF $Z(t)$ is replaced by the average of its diagonal entries, $\tr Z(t)/d$, which is a scalar.%
\footnote{For anisotropic transport, one should work with the covariance tensor $\langle \Delta\vec r(t) \otimes \Delta\vec r(t) \rangle$ instead of the MSD.}

Combining \cref{eq:VACF-overd} with the Green--Kubo type relation for the MSD,
\begin{align}
    \delta r^2(t) =  2 d \int_0^t (t-s) Z(s) \, \diff s\,,
    \label{eq:msd-via-Z(t)}
\end{align}
we arrive at the integral equation
\begin{align}
  \delta r^2(t) = 2d D_0 t - \frac{1}{\zeta_0}\int_0^t  \zeta_\text{reg}(t-s)
  \, \delta r^2(s)\,\diff s\,.
  \label{eq-msd-overd}
\end{align}
The equation follows either from the associativity of the convolution operation or, perhaps more directly, by multiplication of \cref{eq:VACF-overd} with $2d(t-s)$ and integration over $s$ from $0$ to $t$.
At short times, where the integral term is negligible since $\delta r^2(0) = 0$, the asymptote reproduces free Brownian motion,
$\delta r^2(t \to 0) \simeq 2d D_0 t$. As mentioned above, this contribution corresponds to the singular term in the decomposition \eqref{eq:zeta-repres}. Thus, any deviation of the MSD from a linear increase contributes to the memory encapsulated in $\zeta_\text{reg}(t)$.
At long times, one anticipates another diffusive regime,
$\delta r^2(t \to \infty) \simeq 2d D_\infty t$, which defines the effective diffusivity $D_\infty$.
The regular part of the memory function completely encodes a possible enhancement or suppression of $D_\infty$ relative to $D_0$ (see below).

\Cref{eq-msd-overd} is a linear Volterra convolution equation of the second kind with respect to $\delta r^2(t)$ for given $\zeta_\text{reg}(t)$, which may be solved by discretising the integral with, e.g., the trapezoidal rule \cite{Jones:MC1961}. We aim, however, at extracting the memory function $\zeta_\text{reg}(t)$ for a given MSD, $\delta r^2(t)$. For reasons of numerical stability \cite{Straube:CP2020,Kowalik:PRE2019}, we choose to work with the integral of Eq.~(\ref{eq-msd-overd}) to solve for the dimensionless memory integral, $G(t):=\zeta_0^{-1}\int_0^t \zeta_\text{reg}(s) \diff s$, which obeys
\begin{align}
\int_0^t\delta r^2(s) \, \diff s =  d D_0 t^2  - \int_0^t G(t-s) \, \delta r^2(s) \,\diff s
 \label{eq-msd-int-overd}
\end{align}
with the initial condition $G(0)=0$.
Afterwards, the regular part of the memory function follows from differentiation, $\zeta_\text{reg}(t)=\zeta_0 \,G'(t)$.

As a general consequence of the splitting of the friction in \cref{eq:zeta-repres}, the long-time limit of the memory integral, $G_\infty:=G(t\to \infty)$, is related to the effective, macroscopic friction via
\begin{equation}
  \zeta_\infty = \int_0^\infty \zeta(s) \, \diff s = \zeta_0(1 + G_\infty) \,.
\end{equation}
Together with the Green--Kubo and Einstein relations, $\hat Z(\omega \to 0) = D_{\infty} = \kB T/\zeta_\infty$ [see \cref{def:memory}], it follows that $G_\infty$ quantifies the reduction of the diffusivity,
\begin{equation}
 G_\infty = \frac{\Delta D}{D_{\infty}} := \frac{D_0-D_\infty}{D_{\infty}} \,.
 \label{eq:Ginf}
\end{equation}
We emphasise that the relations between $\delta r^2(t)$, $Z(t)$, $\zeta(t)$, and $G(t)$ given in \cref{sec:noneq-GLE,sec:overdamped} are valid in any stationary ensemble and do not hinge on the notion of equilibrium.

\subsection{Numerical estimation of the memory function}
\label{sec:extr-memfun}

In order to estimate the memory function from MSD data, we proceed as follows. First, we assume that $\delta r^2(t)$ is given on a semi-geometrically spaced time grid (see, e.g., \cite{Frenkel:MD,Hoefling:PRL2006,Hoefling:PRL2007,Straube:CP2020}).
Such a grid is a suitable choice to represent the characteristic behaviour of the MSD on logarithmic scales;
moreover, it allows for the efficient sampling of the MSD with a multiple-$\tau$ correlation technique \cite{Frenkel:MD,Colberg:CPC2011}.
A uniform linear grid would become increasingly redundant with increasing lag time $t$ and the use of such a dense grid would unnecessarily introduce a statistical ``high-frequency'' noise in the data at long times, which is detrimental to the numerical stability.
On the other hand, data on a uniform time grid are required for the standard deconvolution algorithm to invert \cref{eq-msd-int-overd}.
For these reasons, we take the data for $\delta r^2(t)$ given on a semi-geometrically spaced grid and interpolate them on a uniform, equidistantly spaced grid, $\delta r^2_i=\delta r^2(t_i)$ with $t_i=i \Delta t$ ($i = 0, 1, \dots $).

Second, given the data points $\delta r^2_i$, we employ the trapezoidal rule to discretise the integrals in \cref{eq-msd-int-overd} and solve for $G_i=G(t_i)$ with the initial values $\delta r^2_0=0$ and $G_0=0$. We arrive at a recurrence relation for $G_i$ ($i>0$):
\begin{align}
    G_i = \frac{1}{\delta r^2_1}\left[dD_0\Delta t(i+1)^2-\sum_{j=1}^i \delta r^2_j
    - \frac{1}{2}\delta r^2_{i+1} - \sum_{j=2}^i G_{i+1-j}\, \delta r^2_{j}\right]. \label{eq:Gi}
\end{align}

Third, having obtained the values $G_i$ on the uniform grid, it remains to compute the memory function, $\zeta_\text{reg}(t)=\zeta_0 \, G'(t)$ using finite differences:
\begin{equation}
  G'_0=(G_1-G_0)/\Delta t \quad \text{and} \quad G'_i=(G_{i+1}-G_{i-1})/(2\Delta t) \quad \text{for} \quad i>0.
\end{equation}
Note the data for $G_i$ may suffer from statistical noise, which results in unphysical oscillations of the derivative, especially at short times, and hence in the estimate of $\zeta_\text{reg}(t)$.
To improve the quality of the results, we did the following. At the interpolation step, we combined a smaller ($\Delta t_1$) and larger ($\Delta t_2$) time resolution, such that $\Delta t_1 < \Delta t < \Delta t_2$. The interpolated data sets for the MSDs yield results for $\zeta_\text{reg}(t)$ which better reproduce shorter and longer times, respectively, and which match on intermediate lag times.

\section{Overdamped Brownian motion under confinement}

In this section, the framework developed in \cref{sec:noneq-GLE,sec:overdamped} is applied to and tested on colloidal diffusion in a sinusoidal potential landscape [\cref{eq:Lang-BP}]. For this basic problem, a number of analytic results have been worked out in the past \cite{Lifson:JCP1962, Festa:PA1978, Stratonovich:1967, Reimann:PRL2001,Reimann:PRE2002, Fulde:PRL1975}.
Due to the symmetry of the potential, it is sufficient to consider only the one-dimensional transport along the $x$-axis, which is normal to the ripples of the potential.

\subsection{Analytic solution for the long-time transport}
\label{sec:analytic-long-time}

In equilibrium ($f=0$), the effective long-time diffusion constant was calculated as \cite{Lifson:JCP1962, Festa:PA1978} % Spiechowicz:E2022
\begin{align}
    D_{\infty} = \frac{D_0}{|I_0(\upsilon)|^2}\,,
    \label{eq:D-inf}
\end{align}
where $\upsilon = U_0 /\kB T$ is the strength of the confinement relative to the thermal energy and $I_{\nu}(x)$ denotes the modified Bessel function of the first kind of order $\nu$. Because of $I_0(0)=1$ and $I_0(x) \ge I_0(0)$, we see that the effective diffusivity is maximum in the absence of the confinement, $D_\infty = D_0$ for $U_0 = 0$, and it is suppressed due to the corrugation of the potential, leading to $D_\infty < D_0$ for $U_0 > 0$.

\begin{figure}
  \includegraphics[width=\linewidth]{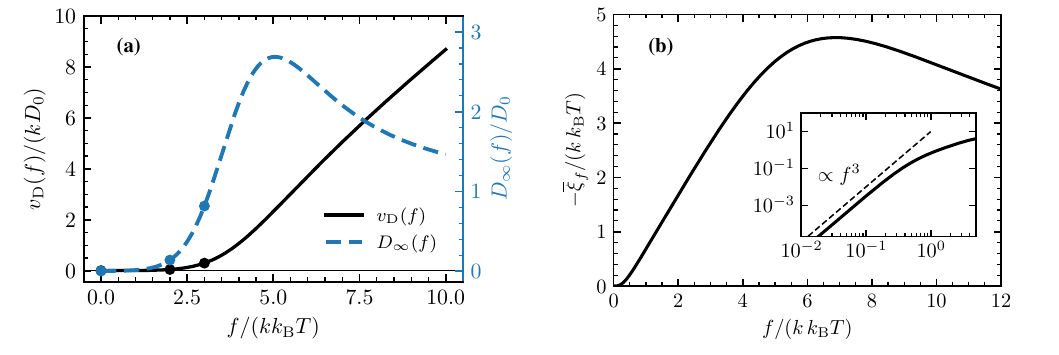}
  \caption{(a) Analytic predictions for the dependence of the long-time transport on the driving force~$f$: mean drift velocity $\vD(f)$ [black solid line, \cref{eq:vD}] and diffusion constant $D_\infty(f)$ [dashed blue line, \cref{eq:Dinf-f}].
  Symbols indicate simulation results (\cref{sec:MSD-numerics}) for exemplary driving forces.
  ~(b)~Non-equilibrium bias $\overline{\xi}_f < 0$ of the random force $\xi_f(t)$ of the GLE [\cref{eq:GLE,eq:GLE-overd}];
  $\overline{\xi}_f$ is calculated from the predictions for $\vD(f)$ and $\zeta_\infty(f) = \kB T / D_\infty(f)$
  using \cref{eq:drift-noise}.
  Inset: double-logarithmic representation of $-\overline{\xi}_f$ in comparison to a behaviour $\sim f^3$ (dashed blue line).
  In both panels, the amplitude of the potential landscape is set to $U_0 = 5 \kB T$.
  }
  \label{fig:longtimes}
\end{figure}

Out of equilibrium ($f>0$) the drift velocity was obtained by \citet{Stratonovich:1967} and reads
\begin{align}
    v_\text{D}(f) = \frac{k D_0}{\pi}\,\frac{\sinh(\pi \kappa)}{|I_{\mathrm{i} \kappa}(\upsilon)|^2}\,,
    \label{eq:vD}
\end{align}
where $\kappa = f/(k \, \kB T)$ specifies the relative driving strength;
the order of the Bessel function becomes purely imaginary and is non-integer valued.
For weak driving, $f \ll k U_0$, the drift $\vD(f)$ is strongly suppressed and varies exponentially, whereas for strong driving, $f \gg k U_0$, it approaches from below the linear increase, $\vD(f) \simeq \zeta_0^{-1} f$,
indicating effectively unconfined transport [\cref{fig:longtimes}(a)].
The corresponding long-time diffusivity was calculated only recently by \citet{Reimann:PRL2001,Reimann:PRE2002}:
\begin{align}
    D_{\infty}(f) = D_0 \frac{ \langle \mathcal{I}_{\pm}(x,f)^2 \, \mathcal{I}_{\mp}(x,f)\rangle_x} {\langle \mathcal{I}_{\pm}(x,f)\rangle_x^3}\,,
    \label{eq:Dinf-f}
\end{align}
in terms of
\begin{equation}
  \mathcal{I}_{\pm}(x,f) = \langle \exp\left[ \pm( U_\text{eff}(x,f)-U_\text{eff}(x\mp y,f) )/\kB T\right]\rangle_y \,;
\end{equation}
$\langle \,\cdots\, \rangle_x = \lambda^{-1} \int_{x_0}^{x_0 + \lambda} \cdots \,\, \diff x$ denotes the spatial average over one wavelength $\lambda$ of the landscape and
$U_\text{eff}(x;f)= - f x + U(x)$ is a tilted potential that, in addition to the landscape, includes the external force.
The dependence of $D_\infty(f)$ on the driving force exhibits a maximum near the critical force $f_\text{cr} = k U_0$ of the athermal ($D_0=0$) depinning transition; for small driving, the diffusivity is strongly suppressed, $D_\infty(f) \ll D_0$ [\cref{fig:longtimes}(a)].
In the linear response regime, the solutions in \cref{eq:vD,eq:Dinf-f} satisfy an Einstein relation:
\begin{equation}
  \vD(f) \simeq \frac{f}{\kB T} \, D_\infty(f) \,, \qquad f \to 0.
\end{equation}

In the context of the non-equilibrium GLE [\cref{eq:GLE,eq:GLE-overd}], the predictions for the long-time transport coefficients allow us to infer the bias $\overline{\xi}_f$ of the random force $\xi_f(t)$.
From inversion of \cref{eq:drift-noise}, it holds
\begin{equation}
  \overline{\xi}_f = [\kB T / D_\infty(f)] \, \vD(f) - f < 0 \,,
\end{equation}
which is negative due to $\vD(f) \leq \zeta_0^{-1} f$ for all $f$.
Below the critical force $f_\text{cr}$, the bias $\overline{\xi}_f$ increases apparently linearly as $f$ is increased [\cref{fig:longtimes}(b)]. However, inspection of the numerical data for weak driving suggests that
$\overline{\xi}_{f\to 0} \sim f^3$ [inset of \cref{fig:longtimes}(b)], which is in line with recent theoretical predictions for the coarse-grained dynamics of a Hamiltonian many-particle system, obtained within the Mori projection formalism \cite{Koch:PRR2024}.

\subsection{Analytic solution for the dynamics in equilibrium}
\label{sec:analytic-sol}

In equilibrium ($f=0$), the underdamped analogue of the dynamic problem in \cref{eq:Lang-BP} was studied by \citet{Fulde:PRL1975}, who give an analytic solution in terms of continued fractions for the frequency-dependent electron mobility,
$\hat{\mu}(\omega):=(\kB T)^{-1}\int_0^t  \e^{\i\omega t} Z(t) \,\diff  t$. They showed that truncating the continued fraction already at the second order provides a reasonably good approximation:
\begin{equation}
  \hat{\mu}(\omega) \approx \left[-\i m \omega + \zeta_0 + \zeta_0 \, \alpha k^2 D_0/(-\i\omega + \tau_\text{m}^{-1})\right]^{-1}
\end{equation}
with constants $\tau_\text{m} > 0$ and $\alpha = \upsilon I_1(\upsilon)/I_0(\upsilon)$.
Performing the overdamped limit, $m \to 0$, in this expression %\cref{eq:mu-fulde}
and proceeding to the VACF, $\hat{Z}(\omega)= \kB T \,\hat{\mu}(\omega)$, we arrive at
\begin{align}
    \hat{Z}(\omega) \approx \cfrac{D_0}{1 + \cfrac{\alpha\tau_0^{-1}}{-\i\omega + \tau_\text{m}^{-1}}} \,,
    \qquad \tau_0 := \frac{1}{k^2 D_0} \,.
    \label{eq:Z(omega)}
\end{align}
All occurring quantities are known except for the time $\tau_\text{m}$, which is found from matching
the long-time diffusivity, $\hat{Z}(\omega \to 0) = D_\infty$, with its value given in \cref{eq:D-inf}, which yields
\begin{align}
    \tau_\text{m} = \frac{ |I_0(\upsilon)|^2-1}{\alpha \tau_0^{-1}}
    = \frac{1}{\alpha}\left(\frac{1}{ k^2 D_\infty} - \frac{1}{ k^2 D_0}\right)
    = \frac{\tau_0}{\alpha} \frac{\Delta D}{D_\infty}\,,
    \label{eq:tau-m2}
\end{align}
where $\Delta D := D_0 -D_\infty \ge 0$. 

The frequency-dependent memory kernel $\hat{\zeta}(\omega)=\zeta_0 + \hat{\zeta}_\text{reg}(\omega)$ follows from the comparison of \cref{eq:Z(omega)} with \cref{def:memory-overd}, which yields for the regular part:
\begin{align}
    \hat{\zeta}_\text{reg}(\omega) \approx \frac{\alpha \tau_0^{-1} \zeta_0}{-\i\omega + \tau_\text{m}^{-1}}\,.
    \label{eq:zeta-reg(omega)}
\end{align}
Transforming this result to the time domain, we obtain the total memory function [\cref{eq:zeta-repres}] with the regular part
\begin{align}
    \zeta_\text{reg}(t) = \zeta_0 \,\alpha \tau_0^{-1}\, \e^{-t/\tau_\text{m}}
        =\zeta_0 \frac{\Delta D}{D_\infty \tau_\text{m}} \e^{-t/\tau_\text{m}} \,,
        \label{eq:zeta-reg}
\end{align}
and the memory integral $G(t)=\zeta_0^{-1} \int_0^t \zeta_\text{reg}(s) \,\diff s$:
\begin{align}
    G(t) = \frac{\Delta D}{D_\infty }\left(1 - \e^{-t/\tau_\text{m}}\right) .
    \label{eq:G-reg}
\end{align}
Hence, the long-time limit, $G_\infty= \Delta D/D_\infty$, quantifies the reduction of the diffusivity [\cref{eq:Ginf}] and $\tau_\text{m}$ is the characteristic timescale of the memory effects in $G(t)$ and thus $\zeta_\text{reg}(t)$.

For the VACF in the time domain, we need to backtransform $\hat{Z}(\omega)$ given in \cref{eq:Z(omega)}. To this end, we single out the Markovian short-time diffusion first:
\begin{align}
    \hat{Z}(\omega) - D_0 = -D_0 \frac{\alpha \tau_0^{-1}}{-\i\omega + \tau_\text{m}^{-1} + \alpha \tau_0^{-1}}\,.
\end{align}
This suggests to define the rate
\begin{equation}
  \tau_\text{c}^{-1} = \tau_\text{m}^{-1} + \alpha \tau_0^{-1}
  \label{eq:tau-c}
\end{equation}
and, thus, we have
\begin{align}
    Z(t) = D_0 \delta_+(t) - \frac{\Delta D}{\tau_\text{c}} \e^{-t/\tau_\text{c}} \,.
    \label{eq:Z(t)}
\end{align}
Here, with account of \cref{eq:tau-m2,eq:tau-c} we made use of the relation $D_0\alpha \tau_0^{-1} = \Delta D \tau_\text{c}^{-1}$.
After further integration, for the time-dependent diffusivity, $D(t)=\int_0^t Z(s) \diff s$, and for the MSD, $\delta x^2(t) = 2 \int_0^t D(s) \diff s = 2 \int_0^t (t-s) Z(s) \diff s$, it follows:
\begin{align}
    D(t) & = D_\infty + \Delta D \, \e^{-t/\tau_\text{c}}, \label{eq:D(t)} \\
    \delta x^2(t) & = 2 D_\infty t + 2\Delta D \,\tau_\text{c}\left( 1 - \e^{-t/\tau_\text{c}}\right). \label{eq:dx2(t)}
\end{align}
These results, relying on the approximation to $\hat Z(\omega)$ [\cref{eq:Z(omega)}] are found to accurately reproduce the true dynamics as obtained from Brownian dynamics simulations (\cref{fig:data}).

We note that the quantities $Z(t)$, $D(t)$ and $\delta x^2(t)$ evolve at the timescale
\begin{align}
    \tau_\text{c} = \frac{D_\infty}{D_0} \tau_\text{m}\,,
    \label{eq:tau-c2}
\end{align}
which means that the memory function varies more slowly, at a stretched timescale, relative to the position/velocity correlations of the particle.
On the other hand, the particle position has two relaxation channels [\cref{eq:tau-c}]: directly via Markovian diffusion, with rate $\alpha\tau_0^{-1}$, and via barrier crossings with rate $\tau_\text{m}^{-1}$.
The latter rate depends on how strongly free diffusion is reduced due to the confinement, which is quantified by $G_\infty = \Delta D / D_\infty$.
In particular for high barriers, $U_0 \gg \kB T$, the mobility of the particle is strongly suppressed, $D_\infty \ll D_0$, which leads to a separation of the timescales at which $Z(t)$, $D(t)$ and $\delta x^2(t)$ at one hand and the memory function $\zeta_\text{reg}(t)$ on the other hand relax: $\tau_\text{c} \ll \tau_\text{m}$.

We note further that the functional forms of \cref{eq:Z(t),eq:D(t),eq:dx2(t)} have already been anticipated in our earlier work \cite{Choudhury:NJP2017}. The main difference is that, there, they were based on an \emph{ad hoc} assumption for the form of $Z(t)$, whereas in this work, they are a consequence of the approximation \cref{eq:Z(omega)}, which includes all parametric dependencies on the confining potential.

\subsection{MSD data from Brownian dynamics simulations} % in and out of equilibrium
\label{sec:MSD-numerics}

We have run Brownian dynamics simulations for the dynamics in \cref{eq:Lang-BP} to obtain data for the MSD at different driving forces $f$.
To this end, we have used $k^{-1}$, $\kB T$, and $\tau_0$ [\cref{eq:Z(omega)}]
as the units of length, energy, and time, respectively.
This choice of scales is equivalent to setting
$k=1$, $\kB T =1$ and $D_0=1$; in particular, it implies $\zeta_0=\kB T/D_0 = 1$.
The remaining control parameters of the problem are the height of the potential barrier, $2 U_0$, and the external force, $f$.
As we have shown in \cref{sec:analytic-sol} for the equilibrium case ($f=0$), the timescales $\tau_\text{c}$, characterising positional correlations (including the MSD, the VACF and the diffusivity) and $\tau_\text{m}$, characterising noise correlations and the memory function of the GLE, may differ strongly. To ensure a visible separation of these timescales on one hand and still to be able to capture both crossovers in the simulations, we set the amplitude of the landscape to $U_0=5\kB T$.
This yields $\tau_\text{c} \approx 0.22\tau_0$ and $\tau_\text{m} \approx 166 \tau_0$.
Going beyond the analytically tractable case $f=0$, we have also simulated non-equilibrium situations with driving forces $f=2 k\, \kB T$ and $f=3 k\, \kB T$ to investigate the effect of driving on the dynamics.

More specifically, we integrated \cref{eq:Lang-BP} numerically up to the time $t=5\times 10^8 \tau_0$ using the Euler--Maruyama scheme with a timestep of $\delta t = 0.0025\tau_0$.
The moments $\langle x(t)^n \rangle$ for $n=1,\dots,4$ were calculated with a moving time average and, additionally, an ensemble average over 30 independent realisations for each value of the force~$f$. The MSD was obtained as the variance
$\delta x^2(t) := \Var(x(t)) = \langle x(t)^2 \rangle - \langle x(t) \rangle ^2$ on a semi-geometric time grid (see \cref{sec:extr-memfun}).

\begin{table*}[b]
% define centred columns of flexible width
\renewcommand\tabularxcolumn[1]{>{\hfill}p{#1}<{\hfill\hbox{}}}
\heavyrulewidth=.1em
\tabcolsep=.25em
\small
% add extra space between the models to separates the cmidrules
\begin{tabularx}{.9\linewidth}{X@{\extracolsep{1.25em}}d{7}d{7}@{\extracolsep{1.25em}}d{7}d{7}@{\extracolsep{1.25em}}d{6}d{6}}
\toprule
% use X-columns to adjust column widths to overall text width,
% centre first part of the table rather than to align on the decimal point
$f/(k\,\kB T)$ &
\multicolumn{2}{c}{$\vD(f) / (k D_0)$} &
\multicolumn{2}{c}{$D_\infty(f) / D_0$} &
\multicolumn{2}{c}{$G_\infty$} \\
\cmidrule{2-3} \cmidrule{4-5} \cmidrule{6-7}
 & \multicolumn{1}{c}{\cref{eq:vD}} & \multicolumn{1}{c}{sim.}
 & \multicolumn{1}{c}{\cref{eq:Dinf-f}} & \multicolumn{1}{c}{sim.}
 & \multicolumn{1}{c}{\cref{eq:Ginf}} & \multicolumn{1}{c}{sim.} \\
\midrule[\heavyrulewidth]
% force   vD   Deff   Ginf
0 & 0 & \multicolumn{1}{c}{$<3\times 10^{-7}$} & 0.001348 & 0.001382 & 740.8 & 740.3 \\
2 & 0.044957 & 0.044959 & 0.13844 & 0.13839 & 6.223 &  6.230 \\
3 & 0.29584  & 0.29577  & 0.81623 & 0.81759 & 0.2252 & 0.2221\\
\bottomrule
\end{tabularx}
\caption{Theoretical predictions and simulation results for the long-time transport coefficients in equilibrium ($f=0$) and far from equilibrium. The columns show the driving force $f$, the drift velocity $\vD(f)$, the effective long-time diffusivity $D_\infty(f)$, and the long-time limit $G_\infty = G(t\to \infty)$ of the dimensionless memory integral.
}
\label{tab:longtimes}
\end{table*}

\begin{figure*}
    \includegraphics[width=0.55\textwidth]{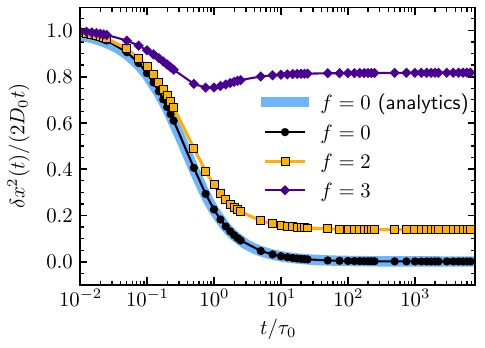}
    \caption{Results for the MSD $\delta x^2(t)$, rescaled by $2D_0 t$ for free diffusion.
    The data were obtained from Brownian dynamics simulations for $U_0=5 \kB T$ at three different values of the driving force $f$.
    The long-time limits of the data correspond the effective diffusivity, $D_{\infty}(f) / D_0$.
    The bold blue line shows the analytic prediction in equilibrium [\cref{eq:dx2(t)}],
    symbols mark actual simulation data, and thin lines are guides to the eye. 
    }
	\label{fig:data}
\end{figure*}

Our approach yields results for the drift velocity $\vD(f)$ and the effective long-time diffusivity $D_\infty(f)$ that agree very well with the corresponding predictions [\cref{eq:vD,eq:Dinf-f}], see \cref{fig:longtimes,tab:longtimes}.

The comparably high potential barrier ($U_0 = 5 \kB T$) considered here leads to a strong suppression of the long-time diffusivity in equilibrium [$D_\infty(f=0)$] by three orders of magnitude relative to the free diffusivity $D_0$.
The MSD shows a crossover between the two values as a function in time (\cref{fig:data}).
To this end, plotting the ratio $\delta x^2(t) / (2D_0t)$ allows one to infer the diffusion coefficients at short and long times directly from the graph;
on double-logarithmic scales, such a graph may be interpreted as a simplistic approximation to the timescale-dependent diffusivity, $D(t) = \partial_t \delta r^2(t) /2$, up to the factor $D_0$.
Moreover, the MSD data for $f=0$ follow closely the approximate analytic solution which is available in equilibrium [\cref{eq:dx2(t)}].
The suppression of $D_\infty(f) / D_0$ is weakened as the driving force is increased from $f=0$ to $f=3 k\,\kB T$
and the monotonic decay of $\delta x^2(t) / (2D_0t)$ becomes non-monotonic for large forces;
for even stronger driving, one anticipates that $D_\infty(f) > D_0$, which is accompanied by a pronounced increase of $\delta x^2(t) / (2D_0t)$ for longer times, $t \gtrsim \tau_0$ (\cref{fig:longtimes}).

\subsection{Results for the memory function and its integral}

\begin{figure*}
    \includegraphics[width=\linewidth]{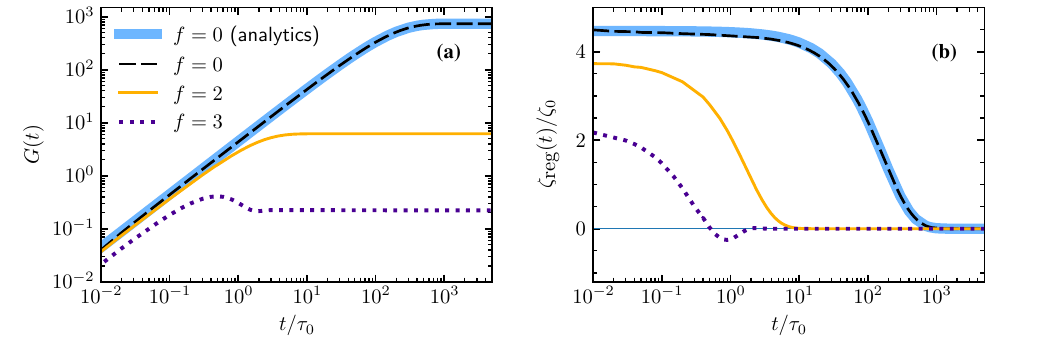}
    \caption{(a) Memory integrals $G(t)$ obtained numerically from the MSD data [\cref{fig:data}] for $U_0=5$ and three different values of the driving force $f$. The bold blue line corresponds to the analytic result in equilibrium ($f=0$) [\cref{eq:G-reg}].
    ~(b)~Memory functions $\zeta_\text{reg}(t) = \zeta_0^{-1} G'(t)$ obtained by numerical differentiation of the data shown in panel~(a). The bold blue line is the prediction in equilibrium [\cref{eq:zeta-reg}].    
    }
	\label{fig:memint-memfun}
\end{figure*}

Based on the MSD data presented in \cref{fig:data}, we have calculated the memory integral $G(t)$ and the regular part of the memory function, $\zeta_\text{reg}(t) = \zeta_0 G'(t)$, as described in \cref{sec:extr-memfun}.
In order to more clearly resolve the behaviour at short and long timescales, we have combined results from data sets with sampling timesteps $\Delta t_1 = 0.001 \tau_0$ and $\Delta t_2 =0.1 \tau_0$ for each force value, which yields overlapping results for $G(t)$ and $\zeta_\text{reg}(t)$ at intermediate timescales.
As a consistency check, the equilibrium data obtained from this approach agree very well with the approximate (yet accurate) analytic solutions, without any fit parameters (\cref{fig:memint-memfun}).

The memory integral $G(t)$ exhibits a linear growth at short times,
$G(t \to 0) \simeq \zeta_0^{-1} \zeta_\text{reg}(t\to 0) \,t$ and saturates monotonically at long times, $G(t\to \infty) = G_\infty$ [\cref{fig:memint-memfun}(a)].
The limiting value satisfies the relation $G_\infty = (D_0 - D_\infty)/D_\infty$ [\cref{eq:Ginf}], also far from equilibrium (\cref{tab:longtimes}).
Due to the high potential barriers considered here ($U_0=5 \kB T$), it holds $D_\infty \ll D_0$ in equilibrium and thus $G_\infty(f=0) \gg 1$.
In equilibrium, the two regimes in $G(t)$ are delimited by the crossover timescale $\tau_\text{m} \approx 166 \tau_0$ for the present model parameters, which is well corroborated by the data in \cref{fig:memint-memfun}(a).
For increasing driving force $f$, the decrease of $D_\infty$ goes hand in hand with a decrease of $G_\infty$ and, concomitantly, of the crossover timescale,
\begin{equation}
 \tau_\text{m} \approx \zeta_0 G_\infty / \zeta_\text{reg}(t\to 0) \,.
 \label{eq:tau-m}
\end{equation}
This relation follows from matching the asymptotic regimes at short and long times;
in equilibrium, it is already implied by the analytic solution in \cref{eq:zeta-reg,eq:G-reg}.
This estimate of the crossover time appears still reasonable for $f=2 k\,\kB T$, but breaks down for larger forces (e.g., $f=3 k\,\kB T$) due to a distinctly non-monotonic behaviour of $G(t)$. In the latter case, $G(t)$ exhibits a maximum at intermediate times (here, at $t\approx 0.5\tau_0$), which hints at the competition of two timescales governing the behaviour of $G(t)$.
We note that, qualitatively speaking, the non-monotonicity of $G(t)$ for $f=3 k\,\kB T$ mirrors the non-monotonic feature of the corresponding rescaled MSD, $\delta r^2(t) / (2 t)$ (\cref{fig:data}).
For stronger driving such that $D_\infty > D_0$, one anticipates from \cref{eq:Ginf} that $G(t)$ becomes negative at long times; in particular, $G_\infty < 0$.

Finally, we have obtained the memory functions $\zeta_\text{reg}(t)$ from numerical differentiation of $G(t)$
for the three investigated values of the driving force [\cref{fig:memint-memfun}(b)].
All curves converge to finite values $\zeta_\text{reg}(t \to 0)$ at short timescales and they decay to zero for long times.
The equilibrium data for $\zeta_\text{reg}(t)$ are in close agreement with the prediction in \cref{eq:zeta-reg},
which exhibits a single-exponential decay to zero with rate $\tau_\text{m}^{-1}$.
Driving the system moderately out of equilibrium, the memory function retains the monotonic decay for $f=2 k\,\kB T$, but it picks up the non-monotonicity in $G(t)$ for the largest force considered here;
in this case, $\zeta_\text{reg}(t)$ shows damped oscillation about zero.
Such a non-monotonic decay of the memory function (and of correlation functions in general) is a genuine non-equilibrium effect that has been observed also in other colloidal systems \cite{Berner:NC2018,Kurzthaler:PRL2018,Kurzthaler:SR2016,Caraglio:PRL2022};
it is excluded for overdamped dynamics in equilibrium, which requires that correlation functions are completely monotone functions
\cite{Feller:Probability:Vol2,Franosch:JPCB1999,Franosch:JSP2002};
only in this case, they may be approximated by a superposition of decaying exponentials with positive weights,
which is also known as a Prony series \cite{Hohenegger:SJAM2018}.

The non-monotonicity of $\zeta_\text{reg}(t)$ for $f=3 k\,\kB T$ is inherited from the time-dependent diffusivity $D(t)$ (see data in \cref{fig:data} for a simplistic approximation).
It results from different physical mechanisms at short and long timescales:
With high probability, the particle starts near a minimum of the potential landscape, which has a similar effect as a harmonic confinement. Thus, the diffusivity at short times decreases from its microscopic value, $D'(t \to 0) < 0$ with $D(0) = D_0$.
At longer times, the finiteness of the effective barrier height becomes relevant, which may speed up the dispersion, in particular, for driving forces near the threshold force $f_\text{cr}$ of the depinning transition.
For weak driving, $D_\infty \ll D_0$ and $D(t)$ decays monotonically to its long-time limit, $D_\infty = D(t \to \infty)$.
For stronger forces such that $D_\infty(f) > D_0$, there is a time window where the diffusivity increases, $D'(t) > 0$, which suggests that the long-time limit is approached from below.
Hence, $D(t)$ is non-monotonic and displays a minimum at intermediate timescales;
the investigated force $f=3 k\,\kB T = 0.6 f_\text{cr}$ is just at the onset of this behaviour.

The short-time value $\zeta_\text{reg}(t \to 0)$ is a non-trivial quantity and is one of the factors determining the persistence time $\tau_\text{m}$ of the memory effects,
which may be defined in general as (see \cref{eq:tau-m} for equilibrium and weak driving)
\begin{equation}
  \tau_\text{m} = \frac{1}{\zeta_\text{reg}(t\to 0)} \int_0^\infty t \,\zeta_\text{reg}(t) \,\diff t \,.
\end{equation}
Overall, one infers from the data shown in \cref{fig:memint-memfun}(b) that the timescale $\tau_\text{m}$ is diminished by three orders of magnitude if the driving is increased from $f=0$ (equilibrium) to $f=3 k\,\kB T$.
We note that $\zeta_\text{reg}(t)$, together with $\zeta_0$, determines the autocorrelation function of the random force $\vec R(t)$, introduced after \cref{eq:GLE-displ}.

\subsection{Non-Gaussian effects}

\begin{figure*}
  \includegraphics[width=\linewidth]{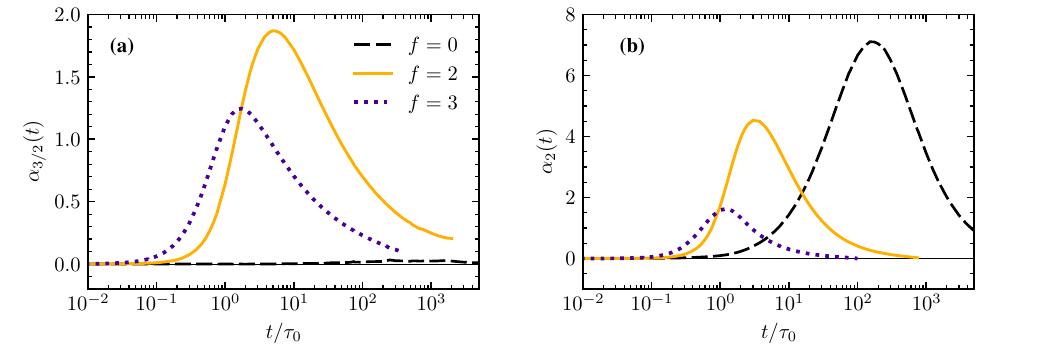}
  \caption{Non-Gaussian parameters of the particle displacement $\Delta x(t)$ after time $t$. The lines show simulation results for the (a) third and (b) fourth normalised cumulants $\alpha_{n/2}(t) := \kappa_{n}(t) / \bigl[c_{n/2} \kappa_{2}(t)^{n/2}\bigr]$ for three different driving forces $f$ (in units of $k\,\kB T$) and fixed potential amplitude $U_0 = 5 \kB T$.
  }
  \label{fig:ngp}
\end{figure*}

The GLEs considered here are linear in their variable, either the velocity in the underdamped description [\cref{eq:GLE}] or the displacement in the overdamped limit [\cref{eq:GLE-displ}].
Combined with a Gaussian (coloured) noise, this implies that the generated solutions for $\vec v(t)$ or $\Delta \vec r(t)$ are Gaussian processes again, since the linear superposition of Gaussian variables (with deterministic coefficients) is Gaussian again \cite{Oksendal:Stochastic}.
In complex transport, this is usually an approximation, the quality of which may be quantified in terms of the so-called non-Gaussian parameters, defined as the normalised cumulants of the displacement \cite{Hoefling:RPP2013,BoonYip:1980}:
\begin{equation}
  \alpha_{n/2}(t) := \frac{\kappa_n(t)}{c_{n/2} \,\kappa_2(t)^{n/2}} \,, \qquad n=3,4,\dots.
\end{equation}
with constants $c_{m} = (2m-1)!! = 2^m \,\Gamma(m+1/2) / \Gamma(1/2)$ % = (2m)!/(2^m m!)
for transport in $d=1$ dimensions, e.g., $c_{3/2} = \sqrt{8/\pi}$ and $c_2 = 3$.
We note that the higher-order cumulants are normalised by $\kappa_2(t) = \delta x^2(t)$, which is the MSD, since the cumulants for odd $n$ vanish for isotropic set-ups without driving.
The cumulants $\kappa_n(t)$ are determined from the characteristic function
$F(q,t) = \langle \exp(\i q \,\Delta x(t))\rangle$ for ``wave number'' $q \to 0$ via
\begin{equation}
  \sum_{n\geq 1} \frac{(\i q)^n}{n!} \, \kappa_n(t) = \log F(q,t) =
  \log \sum_{n\geq 0} \frac{(\i q)^n}{n!} \langle \Delta x(t)^n \rangle \,;
\end{equation}
the second relation allows one to obtain $\kappa_n(t)$ from the moments $\langle \Delta x(t)^{n'} \rangle$ with $n' \leq n$ upon matching powers in $q$.
For Gaussian transport, $F(q,t)$ has a Gaussian shape in $q$ and, thus, the cumulants vanish, $\kappa_n(t)=0$, for $n\geq 3$.
Due to the central limit theorem (CLT), this property is also expected to hold at sufficiently long timescales for transport in a complex environment that appears homogeneous at macroscopic scales \cite{Hoefling:RPP2013}. One shows easily, for independent increments contributing to $\Delta x(t)$, that $\kappa_n(t \to \infty) = O(t)$ and, hence,
$\alpha_{n/2}(t) = O(t^{-n/2+1})$ decays algebraically for long times.

For the present problem of colloidal motion driven across a periodic landscape, we have calculated the non-Gaussian parameters $\alpha_{3/2}(t)$ and $\alpha_2(t)$ from the simulated trajectories for three different driving forces (\cref{fig:ngp}).
In all cases, $\alpha_{n/2}(t \to 0) = 0$ at short times, which reflects the completely overdamped, diffusive motion at short timescales. Also, $\alpha_{n/2}(t \to \infty) = 0$, which follows from the potential landscape being characterised by a single wave length $\lambda = 2\pi/k$; thus, the motion at much larger distances resembles a hopping process of independent steps.
In equilibrium, $\alpha_{3/2}(t)$ remains zero at all times due to symmetry; however, $\alpha_2(t)$ displays a pronounced peak at long times, $t \approx \tau_\text{m}$. This implies that the distribution of $\Delta x(\tau_m)$ deviates considerably from a Gaussian and is much broader than what the corresponding value of the MSD would suggest.

The presence of a driving force $f>0$ diminishes the peak height in $\alpha_2(t)$ as $f$ is increased; at the same time, the peak position shifts to shorter time scales, similarly to what we have observed in the memory function (\cref{fig:memint-memfun}).
The broken isotropy permits a non-trivial behaviour of the odd cumulants: we observe peaks in $\alpha_{3/2}(t)$ at intermediate times which are analogous to the peaks in $\alpha_2(t)$.
In both quantities, the peaks are located at similar $f$-dependent timescales, which are of the same magnitude as the persistence time of the memory effects.

\section{Summary and conclusions}

We have studied memory effects in colloidal motion under confinement and driving. To this end, we have considered a Brownian particle driven over a sinusoidal landscape as a prototypical model system, which also admits experimental implementations \cite{Evstigneev-etal:PRE2008,Straube-Tierno:EPL2013,Juniper-etal:PRE2016,Juniper:NJP2017,Stoop-etal:NL2018}.
Here, we are interested in a GLE where the environment (i.e., the potential landscape) is accounted for by the memory kernel and which reproduces the essential transport behaviour of the full model.
We have addressed two questions:
First, what are the implications of driving from the perspective of coarse-graining, i.e., which modifications of the GLE are necessary to account for driving the dynamics far from equilibrium?
The second question addresses the estimation of memory functions from data, for which established methods exist for the conventional, underdamped GLE. How can one adapt these methods to the case of completely overdamped dynamics, i.e., for instantaneous momentum relaxation?

The starting point was the linear GLE for the particle velocity, amended by a constant, external force [\cref{eq:GLE}].
While it has been clear that the memory function in this approach depends on the external force, it has been recognised only recently \cite{Koch:PRR2024,Shea:SM2024} that, outside of the linear response regime, the random force must be biased [\cref{eq:drift-noise}].
The pronounced non-linear response of the model investigated here appears as an ideal benchmark for this situation, in particular, since detailed analytic predictions are available for the transport properties (\cref{sec:analytic-long-time,sec:analytic-sol}).
Indeed, these results corroborate the previously predicted \cite{Koch:PRR2024} cubic increase of the random force bias, $\overline{\xi}_{f\to 0} \sim f^3$ (\cref{fig:longtimes}).

Considering a non-equilibrium, yet stationary ensemble, it is crucial that autocorrelation functions are defined in terms
of the fluctuations about the mean value; examples for observables with non-vanishing mean are the particle velocity, but also the random force. Given such a correlation function (e.g., the VACF), the analytic properties of the corresponding memory function guarantee the existence of a random force that satisfies a fluctuation--dissipation relation [\cref{eq:coloured-noise}].
Hence, the GLE \eqref{eq:GLE} is justified without resorting to a projected dynamics.
We emphasise that the relations \eqref{def:memory} and \eqref{def:memory-overd} between the VACF and the memory functions are at the heart of this approach to the GLE; they are valid for any stationary relaxation dynamics, in and far from equilibrium.
Nevertheless, it would be desirable and is left for future research to derive \cref{eq:GLE} using, e.g., a Mori projection onto the particle degrees of freedom, but coarse-graining out the potential landscape $U(x)$ (which may be thought of as infinitely many static degrees of freedom).

Regarding the overdamped limit of the GLE, we have decomposed the memory function $\zeta(t)$ into a singular Markovian contribution and a regular part [\cref{eq:zeta-repres}].
This has allowed us to perform the overdamped limit similarly as for the standard Langevin equation
and to obtain a stochastic integral equation for the displacement [\cref{eq:GLE-displ}].
The latter, however, is no longer in the form of an evolution equation;
its solution is driven by a random process $\vec R(t)$ which is a superposition of a free diffusion and
a more regular Gaussian process with its covariance determined by $\zeta_\text{reg}(t)$.
In addition to the advantage of regularisation, the splitting of $\zeta(t)$ collects all memory effects in the regular part.
For the estimation of the memory function from time series data, we have employed an ansatz-free
deconvolution in the time domain; in particular, we have not prescribed a certain form to the memory function.
Following up on our earlier study \cite{Straube:CP2020} and after some adjustments to the case of overdamped dynamics, we have developed a numerical scheme that yields the memory function from MSD data as the sole input (\cref{sec:extr-memfun,fig:memint-memfun}).
From a practical point of view, the non-equilibrium situation is handled by changing to a co-moving coordinate frame, i.e., by subtracting the drift with constant speed $\vD$ from the position data.
For the deconvolution procedure, we suggest to estimate the integral $G(t)$ of the memory function first, which has the trivial initial condition $G(0)=0$ instead of the unknown value of $\zeta_\text{reg}(t\to 0)$ that is needed when solving for $\zeta(t)$ directly. The memory function is eventually obtained from numerical differentiation of $G(t)$.
We have demonstrated the high accuracy of the approach by comparing the estimated long-time limits $G_\infty = G(t\to\infty)$ to the analytic predictions (\cref{tab:longtimes}).
Moreover, we have validated the full time dependence of the estimated memory function against accurate predictions in equilibrium, without any parameter fitting [\cref{fig:memint-memfun,eq:zeta-reg}].

Our analysis of the equilibrium dynamics, which is based on a continued-fraction representation of the frequency-dependent mobility \cite{Fulde:PRL1975}, shows that the relaxation of positional correlations (encoded in quantities such as the MSD, the VACF and the timescale-dependent diffusivity) is governed by the timescale $\tau_\text{c}$, which is different from the decay time $\tau_\text{m}$ of the memory function.
For a large amplitude $U_0$ of the potential landscape, the long-time diffusion of the Brownian particle is strongly suppressed, $D_\infty \ll D_0$, which implies a clear separation of the two timescales, $\tau_\text{c} \ll \tau_\text{m}$ [\cref{eq:tau-c}].
Thus, memory effects can still be present at timescales orders of magnitude longer than what the convergence of the MSD to its long-time asymptote would suggest (\cref{fig:data}).
At variance to this observation, the memory relaxation time is the relevant timescale for the non-Gaussian parameters, which exhibit pronounced peaks near $\tau_\text{m}$ (\cref{fig:ngp}).
The comparably fast convergence of the MSD can be understood from the equilibrium result
$\tau_\text{c}^{-1} = \tau_\text{m}^{-1} + \alpha \tau_0^{-1}$ [\cref{eq:tau-c}]: the superposition of rates suggests
that Markovian diffusion ($\alpha \tau_0^{-1}$) is another relaxation mechanism for the Gaussian part of positional correlations, in addition to the much slower barrier crossing ($\tau_\text{m}^{-1}$).

Out of equilibrium, our study relies on data from Brownian dynamics simulations.
We have exemplified the implications of steady driving for the transport behaviour and its memory function using two intermediate values of the driving force, which are already deep in the non-linear response regime but still well below the depinning transition.
The driving results in a systematic reduction of the memory with respect to its magnitude $\zeta_\text{reg}(t\to 0)$ and its persistence time $\tau_\text{m}$;
in particular, the total integral of the memory function $G_\infty = \Delta D/D_\infty$ decreases.
These trends are expected to change for stronger driving such that $\Delta D < 0 $, which implies $G_\infty < 0$ and thus requires an extended temporal regime where $\zeta_\text{reg}(t) < 0$.
The onset of such a behaviour is already anticipated from the non-monotonic variation of $\zeta(t)$ for the larger of the two driving forces ($f=3k\,\kB T$).
For overdamped dynamics, such a non-monotonic decay of the memory function is a unique fingerprint of non-equilibrium; similar observations have been made for other colloidal systems out of equilibrium \cite{Berner:NC2018, Kurzthaler:PRL2018, Kurzthaler:SR2016, Caraglio:PRL2022}.
The non-monotonic behaviour of $\zeta(t)$ is inherited from the timescale-dependent diffusivity $D(t)$ and we have identified the underlying physical mechanism as a competition between confined motion (decreasing the diffusivity at short times) and depinning (which, for sufficiently strong driving, leads to an enhanced dispersion at long times).

In summary, we have put forward a non-equilibrium GLE for the overdamped dynamics of colloidal motion in an external potential;
as a consequence of driving the dynamics out of the linear response regime, the random force entering the GLE has a non-vanishing mean.
Under the Gaussian approximation for the random force, the coarse-grained dynamics reproduces the transport properties and the dynamic responses correctly at the level of the first and second moments of the displacements (including corresponding correlation functions such as the VACF).
Our simulation results, however, indicate pronounced deviations from a Gaussian displacement distribution (\cref{fig:ngp}), which requires that the random force is a non-Gaussian process.
To this end, it would be of interest to characterise the statistics and the degree of the non-Gaussianity of the random force $\vec \xi_f(t)$ or, equivalently, of the random displacements $\vec R(t)$ in \cref{eq:GLE-displ}, obtained within Brownian dynamics simulations.

The present study was restricted to driving forces below the critical depinning force $f_\text{cr}$. It appears as a rewarding future task to investigate the ramifications of the depinning transition on the memory function and to test for non-analytic anomalies,
in particular, in the low temperature limit, where $D_\infty(f_\text{cr})$ diverges due to the phenomenon of giant diffusion.
Eventually, another interesting variation of the present problem is obtained by utilising self-propulsion as a means to drive the colloidal dynamics out of equilibrium \cite{Straube:2023}.
Can the additional internal degrees of freedom (i.e., the orientation of an anisotropic particle) be incorporated into the memory function of the GLE? How would such a coarse-grained dynamics respond to an external driving force?

\begin{acknowledgments}
 We thank Hugues Meyer and Friederike Schmid for helpful discussions and for pointing us to some new literature (Refs.~\cite{Koch:PRR2024,Shea:SM2024}).
 Financial support by Deutsche Forschungsgemeinschaft (DFG, German Research Foundation)
 %  under Germany’s Excellence Strategy---MATH+: The Berlin Mathematics Research Center (EXC-2046/1)---Project No.\ 390685689 (subproject AA1-18) and further
 under Project No.\ 523950429 is gratefully acknowledged.
\end{acknowledgments}

\bibliography{manuscript}

%aipnum4-2.bst 2019-01-14 (MD) hand-edited version of apsrev4-1.bst
%Control: production of eprint (1) enabled
\begin{thebibliography}{78}%
\makeatletter
\providecommand \@ifxundefined [1]{%
 \@ifx{#1\undefined}
}%
\providecommand \@ifnum [1]{%
 \ifnum #1\expandafter \@firstoftwo
 \else \expandafter \@secondoftwo
 \fi
}%
\providecommand \@ifx [1]{%
 \ifx #1\expandafter \@firstoftwo
 \else \expandafter \@secondoftwo
 \fi
}%
\providecommand \natexlab [1]{#1}%
\providecommand \enquote  [1]{``#1''}%
\providecommand \bibnamefont  [1]{#1}%
\providecommand \bibfnamefont [1]{#1}%
\providecommand \citenamefont [1]{#1}%
\providecommand \href@noop [0]{\@secondoftwo}%
\providecommand \href [0]{\begingroup \@sanitize@url \@href}%
\providecommand \@href[1]{\@@startlink{#1}\@@href}%
\providecommand \@@href[1]{\endgroup#1\@@endlink}%
\providecommand \@sanitize@url [0]{\catcode `\\12\catcode `\$12\catcode
  `\&12\catcode `\#12\catcode `\^12\catcode `\_12\catcode `\%12\relax}%
\providecommand \@@startlink[1]{}%
\providecommand \@@endlink[0]{}%
\providecommand \url  [0]{\begingroup\@sanitize@url \@url }%
\providecommand \@url [1]{\endgroup\@href {#1}{\urlprefix }}%
\providecommand \urlprefix  [0]{URL }%
\providecommand \Eprint [0]{\href }%
\providecommand \doibase [0]{https://doi.org/}%
\providecommand \selectlanguage [0]{\@gobble}%
\providecommand \bibinfo  [0]{\@secondoftwo}%
\providecommand \bibfield  [0]{\@secondoftwo}%
\providecommand \translation [1]{[#1]}%
\providecommand \BibitemOpen [0]{}%
\providecommand \bibitemStop [0]{}%
\providecommand \bibitemNoStop [0]{.\EOS\space}%
\providecommand \EOS [0]{\spacefactor3000\relax}%
\providecommand \BibitemShut  [1]{\csname bibitem#1\endcsname}%
\let\auto@bib@innerbib\@empty
%</preamble>
\bibitem [{\citenamefont {Hansen}\ and\ \citenamefont
  {McDonald}(2006)}]{Hansen:SimpleLiquids}%
  \BibitemOpen
  \bibfield  {author} {\bibinfo {author} {\bibfnamefont {J.-P.}\ \bibnamefont
  {Hansen}}\ and\ \bibinfo {author} {\bibfnamefont {I.}~\bibnamefont
  {McDonald}},\ }\href@noop {} {\emph {\bibinfo {title} {Theory of Simple
  Liquids}}},\ \bibinfo {edition} {3rd}\ ed.\ (\bibinfo  {publisher} {Academic
  Press},\ \bibinfo {address} {Amsterdam},\ \bibinfo {year} {2006})\BibitemShut
  {NoStop}%
\bibitem [{\citenamefont {Boon}\ and\ \citenamefont
  {Yip}(1991{\natexlab{a}})}]{BoonYip:Molecular}%
  \BibitemOpen
  \bibfield  {author} {\bibinfo {author} {\bibfnamefont {J.~P.}\ \bibnamefont
  {Boon}}\ and\ \bibinfo {author} {\bibfnamefont {S.}~\bibnamefont {Yip}},\
  }\href@noop {} {\emph {\bibinfo {title} {Molecular Hydrodynamics}}}\
  (\bibinfo  {publisher} {Dover Publications, Inc.},\ \bibinfo {address} {New
  York},\ \bibinfo {year} {1991})\ \bibinfo {note} {reprint}\BibitemShut
  {NoStop}%
\bibitem [{\citenamefont {Höf{}ling}\ and\ \citenamefont
  {Franosch}(2013)}]{Hoefling:RPP2013}%
  \BibitemOpen
  \bibfield  {author} {\bibinfo {author} {\bibfnamefont {F.}~\bibnamefont
  {Höf{}ling}}\ and\ \bibinfo {author} {\bibfnamefont {T.}~\bibnamefont
  {Franosch}},\ }\bibfield  {title} {\enquote {\bibinfo {title} {Anomalous
  transport in the crowded world of biological cells},}\ }\href
  {https://doi.org/10.1088/0034-4885/76/4/046602} {\bibfield  {journal}
  {\bibinfo  {journal} {Rep. Prog. Phys.}\ }\textbf {\bibinfo {volume} {76}},\
  \bibinfo {pages} {046602} (\bibinfo {year} {2013})}\BibitemShut {NoStop}%
\bibitem [{\citenamefont {Puertas}\ and\ \citenamefont
  {Voigtmann}(2014)}]{Puertas:JPCM2014}%
  \BibitemOpen
  \bibfield  {author} {\bibinfo {author} {\bibfnamefont {A.~M.}\ \bibnamefont
  {Puertas}}\ and\ \bibinfo {author} {\bibfnamefont {T.}~\bibnamefont
  {Voigtmann}},\ }\bibfield  {title} {\enquote {\bibinfo {title} {Microrheology
  of colloidal systems},}\ }\href
  {https://doi.org/10.1088/0953-8984/26/24/243101} {\bibfield  {journal}
  {\bibinfo  {journal} {J. Phys.: Condens. Matter}\ }\textbf {\bibinfo {volume}
  {26}},\ \bibinfo {pages} {243101} (\bibinfo {year} {2014})}\BibitemShut
  {NoStop}%
\bibitem [{\citenamefont {Waigh}(2016)}]{Waigh:RPP2016}%
  \BibitemOpen
  \bibfield  {author} {\bibinfo {author} {\bibfnamefont {T.~A.}\ \bibnamefont
  {Waigh}},\ }\bibfield  {title} {\enquote {\bibinfo {title} {Advances in the
  microrheology of complex fluids},}\ }\href
  {https://doi.org/10.1088/0034-4885/79/7/074601} {\bibfield  {journal}
  {\bibinfo  {journal} {Rep. Prog. Phys.}\ }\textbf {\bibinfo {volume} {79}},\
  \bibinfo {pages} {074601} (\bibinfo {year} {2016})}\BibitemShut {NoStop}%
\bibitem [{\citenamefont {Rigato}\ \emph {et~al.}(2017)\citenamefont {Rigato},
  \citenamefont {Miyagi}, \citenamefont {Scheuring},\ and\ \citenamefont
  {Rico}}]{Rigato:NP2017}%
  \BibitemOpen
  \bibfield  {author} {\bibinfo {author} {\bibfnamefont {A.}~\bibnamefont
  {Rigato}}, \bibinfo {author} {\bibfnamefont {A.}~\bibnamefont {Miyagi}},
  \bibinfo {author} {\bibfnamefont {S.}~\bibnamefont {Scheuring}},\ and\
  \bibinfo {author} {\bibfnamefont {F.}~\bibnamefont {Rico}},\ }\bibfield
  {title} {\enquote {\bibinfo {title} {High-frequency microrheology reveals
  cytoskeleton dynamics in living cells},}\ }\href
  {https://doi.org/10.1038/nphys4104} {\bibfield  {journal} {\bibinfo
  {journal} {Nat. Phys.}\ }\textbf {\bibinfo {volume} {13}},\ \bibinfo {pages}
  {771} (\bibinfo {year} {2017})}\BibitemShut {NoStop}%
\bibitem [{\citenamefont {Zwanzig}(1960)}]{Zwanzig:1960}%
  \BibitemOpen
  \bibfield  {author} {\bibinfo {author} {\bibfnamefont {R.}~\bibnamefont
  {Zwanzig}},\ }\bibfield  {title} {\enquote {\bibinfo {title} {Ensemble method
  in the theory of irreversibility},}\ }\href
  {https://doi.org/10.1063/1.1731409} {\bibfield  {journal} {\bibinfo
  {journal} {J. Chem. Phys.}\ }\textbf {\bibinfo {volume} {33}},\ \bibinfo
  {pages} {1338} (\bibinfo {year} {1960})}\BibitemShut {NoStop}%
\bibitem [{\citenamefont {Mori}(1965)}]{Mori:1965}%
  \BibitemOpen
  \bibfield  {author} {\bibinfo {author} {\bibfnamefont {H.}~\bibnamefont
  {Mori}},\ }\bibfield  {title} {\enquote {\bibinfo {title} {Transport,
  collective motion, and {B}rownian motion},}\ }\href
  {https://doi.org/10.1143/ptp.33.423} {\bibfield  {journal} {\bibinfo
  {journal} {Prog. Theor. Phys.}\ }\textbf {\bibinfo {volume} {33}},\ \bibinfo
  {pages} {423} (\bibinfo {year} {1965})}\BibitemShut {NoStop}%
\bibitem [{\citenamefont {Kubo}(1966)}]{Kubo:RPP1966}%
  \BibitemOpen
  \bibfield  {author} {\bibinfo {author} {\bibfnamefont {R.}~\bibnamefont
  {Kubo}},\ }\bibfield  {title} {\enquote {\bibinfo {title} {The
  fluctuation-dissipation theorem},}\ }\href
  {https://doi.org/10.1088/0034-4885/29/1/306} {\bibfield  {journal} {\bibinfo
  {journal} {Rep. Prog. Phys.}\ }\textbf {\bibinfo {volume} {29}},\ \bibinfo
  {pages} {255} (\bibinfo {year} {1966})}\BibitemShut {NoStop}%
\bibitem [{\citenamefont {Schilling}(2022)}]{Schilling:PR2022}%
  \BibitemOpen
  \bibfield  {author} {\bibinfo {author} {\bibfnamefont {T.}~\bibnamefont
  {Schilling}},\ }\bibfield  {title} {\enquote {\bibinfo {title}
  {Coarse-grained modelling out of equilibrium},}\ }\href
  {https://doi.org/10.1016/j.physrep.2022.04.006} {\bibfield  {journal}
  {\bibinfo  {journal} {Phys. Rep.}\ }\textbf {\bibinfo {volume} {972}},\
  \bibinfo {pages} {1} (\bibinfo {year} {2022})}\BibitemShut {NoStop}%
\bibitem [{\citenamefont {Shea}, \citenamefont {Jung},\ and\ \citenamefont
  {Schmid}(2024)}]{Shea:SM2024}%
  \BibitemOpen
  \bibfield  {author} {\bibinfo {author} {\bibfnamefont {J.}~\bibnamefont
  {Shea}}, \bibinfo {author} {\bibfnamefont {G.}~\bibnamefont {Jung}},\ and\
  \bibinfo {author} {\bibfnamefont {F.}~\bibnamefont {Schmid}},\ }\bibfield
  {title} {\enquote {\bibinfo {title} {Force renormalization for probes
  immersed in an active bath},}\ }\href {https://doi.org/10.1039/d3sm01387a}
  {\bibfield  {journal} {\bibinfo  {journal} {Soft Matter}\ }\textbf {\bibinfo
  {volume} {20}},\ \bibinfo {pages} {1767} (\bibinfo {year}
  {2024})}\BibitemShut {NoStop}%
\bibitem [{\citenamefont {Milster}\ \emph {et~al.}(2024)\citenamefont
  {Milster}, \citenamefont {Koch}, \citenamefont {Widder}, \citenamefont
  {Schilling},\ and\ \citenamefont {Dzubiella}}]{Milster:JCP2024}%
  \BibitemOpen
  \bibfield  {author} {\bibinfo {author} {\bibfnamefont {S.}~\bibnamefont
  {Milster}}, \bibinfo {author} {\bibfnamefont {F.}~\bibnamefont {Koch}},
  \bibinfo {author} {\bibfnamefont {C.}~\bibnamefont {Widder}}, \bibinfo
  {author} {\bibfnamefont {T.}~\bibnamefont {Schilling}},\ and\ \bibinfo
  {author} {\bibfnamefont {J.}~\bibnamefont {Dzubiella}},\ }\bibfield  {title}
  {\enquote {\bibinfo {title} {Tracer dynamics in polymer networks:
  {G}eneralized {L}angevin description},}\ }\href
  {https://doi.org/10.1063/5.0189166} {\bibfield  {journal} {\bibinfo
  {journal} {J. Chem. Phys.}\ }\textbf {\bibinfo {volume} {160}},\ \bibinfo
  {pages} {094901} (\bibinfo {year} {2024})}\BibitemShut {NoStop}%
\bibitem [{\citenamefont {Glatzel}\ and\ \citenamefont
  {Schilling}(2021)}]{Glatzel:EPL2021}%
  \BibitemOpen
  \bibfield  {author} {\bibinfo {author} {\bibfnamefont {F.}~\bibnamefont
  {Glatzel}}\ and\ \bibinfo {author} {\bibfnamefont {T.}~\bibnamefont
  {Schilling}},\ }\bibfield  {title} {\enquote {\bibinfo {title} {The interplay
  between memory and potentials of mean force: {A} discussion on the structure
  of equations of motion for coarse-grained observables},}\ }\href
  {https://doi.org/10.1209/0295-5075/ac35ba} {\bibfield  {journal} {\bibinfo
  {journal} {Europhys. Lett.}\ }\textbf {\bibinfo {volume} {136}},\ \bibinfo
  {pages} {36001} (\bibinfo {year} {2021})}\BibitemShut {NoStop}%
\bibitem [{\citenamefont {Ayaz}\ \emph {et~al.}(2022)\citenamefont {Ayaz},
  \citenamefont {Scalfi}, \citenamefont {Dalton},\ and\ \citenamefont
  {Netz}}]{Ayaz:PRE2022}%
  \BibitemOpen
  \bibfield  {author} {\bibinfo {author} {\bibfnamefont {C.}~\bibnamefont
  {Ayaz}}, \bibinfo {author} {\bibfnamefont {L.}~\bibnamefont {Scalfi}},
  \bibinfo {author} {\bibfnamefont {B.~A.}\ \bibnamefont {Dalton}},\ and\
  \bibinfo {author} {\bibfnamefont {R.~R.}\ \bibnamefont {Netz}},\ }\bibfield
  {title} {\enquote {\bibinfo {title} {Generalized {L}angevin equation with a
  nonlinear potential of mean force and nonlinear memory friction from a hybrid
  projection scheme},}\ }\href {https://doi.org/10.1103/physreve.105.054138}
  {\bibfield  {journal} {\bibinfo  {journal} {Phys. Rev. E}\ }\textbf {\bibinfo
  {volume} {105}},\ \bibinfo {pages} {054138} (\bibinfo {year}
  {2022})}\BibitemShut {NoStop}%
\bibitem [{\citenamefont {Vroylandt}(2022)}]{Vroylandt:EPL2022}%
  \BibitemOpen
  \bibfield  {author} {\bibinfo {author} {\bibfnamefont {H.}~\bibnamefont
  {Vroylandt}},\ }\bibfield  {title} {\enquote {\bibinfo {title} {On the
  derivation of the generalized {L}angevin equation and the
  fluctuation-dissipation theorem},}\ }\href
  {https://doi.org/10.1209/0295-5075/acab7d} {\bibfield  {journal} {\bibinfo
  {journal} {EPL (Europhys. Lett.)}\ }\textbf {\bibinfo {volume} {140}},\
  \bibinfo {pages} {62003} (\bibinfo {year} {2022})}\BibitemShut {NoStop}%
\bibitem [{\citenamefont {Di~Cairano}(2022)}]{DiCairano:JPC2022}%
  \BibitemOpen
  \bibfield  {author} {\bibinfo {author} {\bibfnamefont {L.}~\bibnamefont
  {Di~Cairano}},\ }\bibfield  {title} {\enquote {\bibinfo {title} {On the
  derivation of a nonlinear generalized {L}angevin equation},}\ }\href
  {https://doi.org/10.1088/2399-6528/ac438d} {\bibfield  {journal} {\bibinfo
  {journal} {J. Phys. Commun.}\ }\textbf {\bibinfo {volume} {6}},\ \bibinfo
  {pages} {015002} (\bibinfo {year} {2022})}\BibitemShut {NoStop}%
\bibitem [{\citenamefont {Meyer}, \citenamefont {Voigtmann},\ and\
  \citenamefont {Schilling}(2017)}]{Meyer:JCP2017}%
  \BibitemOpen
  \bibfield  {author} {\bibinfo {author} {\bibfnamefont {H.}~\bibnamefont
  {Meyer}}, \bibinfo {author} {\bibfnamefont {T.}~\bibnamefont {Voigtmann}},\
  and\ \bibinfo {author} {\bibfnamefont {T.}~\bibnamefont {Schilling}},\
  }\bibfield  {title} {\enquote {\bibinfo {title} {On the non-stationary
  generalized {L}angevin equation},}\ }\href
  {https://doi.org/10.1063/1.5006980} {\bibfield  {journal} {\bibinfo
  {journal} {J. Chem. Phys.}\ }\textbf {\bibinfo {volume} {147}},\ \bibinfo
  {pages} {214110} (\bibinfo {year} {2017})}\BibitemShut {NoStop}%
\bibitem [{\citenamefont {Netz}(2018)}]{Netz:JCP2018}%
  \BibitemOpen
  \bibfield  {author} {\bibinfo {author} {\bibfnamefont {R.~R.}\ \bibnamefont
  {Netz}},\ }\bibfield  {title} {\enquote {\bibinfo {title}
  {Fluctuation-dissipation relation and stationary distribution of an exactly
  solvable many-particle model for active biomatter far from equilibrium},}\
  }\href {https://doi.org/10.1063/1.5020654} {\bibfield  {journal} {\bibinfo
  {journal} {J. Chem. Phys.}\ }\textbf {\bibinfo {volume} {148}},\ \bibinfo
  {pages} {185101} (\bibinfo {year} {2018})}\BibitemShut {NoStop}%
\bibitem [{\citenamefont {Jung}\ and\ \citenamefont
  {Schmid}(2021)}]{Jung:SM2021}%
  \BibitemOpen
  \bibfield  {author} {\bibinfo {author} {\bibfnamefont {G.}~\bibnamefont
  {Jung}}\ and\ \bibinfo {author} {\bibfnamefont {F.}~\bibnamefont {Schmid}},\
  }\bibfield  {title} {\enquote {\bibinfo {title} {Fluctuation–dissipation
  relations far from equilibrium: a case study},}\ }\href
  {https://doi.org/10.1039/d1sm00521a} {\bibfield  {journal} {\bibinfo
  {journal} {Soft Matter}\ }\textbf {\bibinfo {volume} {17}},\ \bibinfo {pages}
  {6413} (\bibinfo {year} {2021})}\BibitemShut {NoStop}%
\bibitem [{\citenamefont {Doerries}, \citenamefont {Loos},\ and\ \citenamefont
  {Klapp}(2021)}]{Doerries:JSM2021}%
  \BibitemOpen
  \bibfield  {author} {\bibinfo {author} {\bibfnamefont {T.~J.}\ \bibnamefont
  {Doerries}}, \bibinfo {author} {\bibfnamefont {S.~A.~M.}\ \bibnamefont
  {Loos}},\ and\ \bibinfo {author} {\bibfnamefont {S.~H.~L.}\ \bibnamefont
  {Klapp}},\ }\bibfield  {title} {\enquote {\bibinfo {title} {Correlation
  functions of non-{M}arkovian systems out of equilibrium: analytical
  expressions beyond single-exponential memory},}\ }\href
  {https://doi.org/10.1088/1742-5468/abdead} {\bibfield  {journal} {\bibinfo
  {journal} {J. Stat. Mech.}\ }\textbf {\bibinfo {volume} {2021}},\ \bibinfo
  {pages} {033202} (\bibinfo {year} {2021})}\BibitemShut {NoStop}%
\bibitem [{\citenamefont {Koch}, \citenamefont {Erle},\ and\ \citenamefont
  {Schilling}(2024)}]{Koch:PRR2024}%
  \BibitemOpen
  \bibfield  {author} {\bibinfo {author} {\bibfnamefont {F.}~\bibnamefont
  {Koch}}, \bibinfo {author} {\bibfnamefont {J.}~\bibnamefont {Erle}},\ and\
  \bibinfo {author} {\bibfnamefont {T.}~\bibnamefont {Schilling}},\ }\bibfield
  {title} {\enquote {\bibinfo {title} {Nonequilibrium solvent response force:
  {W}hat happens if you push a {B}rownian particle},}\ }\href
  {https://doi.org/10.1103/physrevresearch.6.l012032} {\bibfield  {journal}
  {\bibinfo  {journal} {Phys. Rev. Research}\ }\textbf {\bibinfo {volume}
  {6}},\ \bibinfo {pages} {l012032} (\bibinfo {year} {2024})}\BibitemShut
  {NoStop}%
\bibitem [{\citenamefont {Jung}(2023)}]{Jung:2023}%
  \BibitemOpen
  \bibfield  {author} {\bibinfo {author} {\bibfnamefont {G.}~\bibnamefont
  {Jung}},\ }\href@noop {} {\enquote {\bibinfo {title} {Mobility, response and
  transport in non-equilibrium coarse-grained models},}\ } (\bibinfo {year}
  {2023})\BibitemShut {NoStop}%
\bibitem [{\citenamefont {Gottwald}\ \emph {et~al.}(2015)\citenamefont
  {Gottwald}, \citenamefont {Karsten}, \citenamefont {Ivanov},\ and\
  \citenamefont {Kühn}}]{Gottwald:JCP2015}%
  \BibitemOpen
  \bibfield  {author} {\bibinfo {author} {\bibfnamefont {F.}~\bibnamefont
  {Gottwald}}, \bibinfo {author} {\bibfnamefont {S.}~\bibnamefont {Karsten}},
  \bibinfo {author} {\bibfnamefont {S.~D.}\ \bibnamefont {Ivanov}},\ and\
  \bibinfo {author} {\bibfnamefont {O.}~\bibnamefont {Kühn}},\ }\bibfield
  {title} {\enquote {\bibinfo {title} {Parametrizing linear generalized
  {L}angevin dynamics from explicit molecular dynamics simulations},}\ }\href
  {https://doi.org/10.1063/1.4922941} {\bibfield  {journal} {\bibinfo
  {journal} {J. Chem. Phys.}\ }\textbf {\bibinfo {volume} {142}},\ \bibinfo
  {pages} {244110} (\bibinfo {year} {2015})}\BibitemShut {NoStop}%
\bibitem [{\citenamefont {Daldrop}, \citenamefont {Kowalik},\ and\
  \citenamefont {Netz}(2017)}]{Daldrop:PRX2017}%
  \BibitemOpen
  \bibfield  {author} {\bibinfo {author} {\bibfnamefont {J.~O.}\ \bibnamefont
  {Daldrop}}, \bibinfo {author} {\bibfnamefont {B.~G.}\ \bibnamefont
  {Kowalik}},\ and\ \bibinfo {author} {\bibfnamefont {R.~R.}\ \bibnamefont
  {Netz}},\ }\bibfield  {title} {\enquote {\bibinfo {title} {External potential
  modifies friction of molecular solutes in water},}\ }\href
  {https://doi.org/10.1103/PhysRevX.7.041065} {\bibfield  {journal} {\bibinfo
  {journal} {Phys. Rev. X}\ }\textbf {\bibinfo {volume} {7}},\ \bibinfo {pages}
  {041065} (\bibinfo {year} {2017})}\BibitemShut {NoStop}%
\bibitem [{\citenamefont {McKinley}\ and\ \citenamefont
  {Nguyen}(2018)}]{McKinley:SJMA2018}%
  \BibitemOpen
  \bibfield  {author} {\bibinfo {author} {\bibfnamefont {S.~A.}\ \bibnamefont
  {McKinley}}\ and\ \bibinfo {author} {\bibfnamefont {H.~D.}\ \bibnamefont
  {Nguyen}},\ }\bibfield  {title} {\enquote {\bibinfo {title} {Anomalous
  diffusion and the generalized {L}angevin equation},}\ }\href
  {https://doi.org/10.1137/17M115517X} {\bibfield  {journal} {\bibinfo
  {journal} {SIAM J. Math. Anal.}\ }\textbf {\bibinfo {volume} {50}},\ \bibinfo
  {pages} {5119 } (\bibinfo {year} {2018})}\BibitemShut {NoStop}%
\bibitem [{\citenamefont {Jung}, \citenamefont {Hanke},\ and\ \citenamefont
  {Schmid}(2017)}]{Jung:JCTC2017}%
  \BibitemOpen
  \bibfield  {author} {\bibinfo {author} {\bibfnamefont {G.}~\bibnamefont
  {Jung}}, \bibinfo {author} {\bibfnamefont {M.}~\bibnamefont {Hanke}},\ and\
  \bibinfo {author} {\bibfnamefont {F.}~\bibnamefont {Schmid}},\ }\bibfield
  {title} {\enquote {\bibinfo {title} {Iterative reconstruction of memory
  kernels},}\ }\href {https://doi.org/10.1021/acs.jctc.7b00274} {\bibfield
  {journal} {\bibinfo  {journal} {J. Chem. Theory Comput.}\ }\textbf {\bibinfo
  {volume} {13}},\ \bibinfo {pages} {2481} (\bibinfo {year}
  {2017})}\BibitemShut {NoStop}%
\bibitem [{\citenamefont {Meyer}, \citenamefont {Pelagejcev},\ and\
  \citenamefont {Schilling}(2020)}]{Meyer:EPL2020}%
  \BibitemOpen
  \bibfield  {author} {\bibinfo {author} {\bibfnamefont {H.}~\bibnamefont
  {Meyer}}, \bibinfo {author} {\bibfnamefont {P.}~\bibnamefont {Pelagejcev}},\
  and\ \bibinfo {author} {\bibfnamefont {T.}~\bibnamefont {Schilling}},\
  }\bibfield  {title} {\enquote {\bibinfo {title} {Non-{M}arkovian
  out-of-equilibrium dynamics: {A} general numerical procedure to construct
  time-dependent memory kernels for coarse-grained observables},}\ }\href
  {https://doi.org/10.1209/0295-5075/128/40001} {\bibfield  {journal} {\bibinfo
   {journal} {EPL (Europhys. Lett.)}\ }\textbf {\bibinfo {volume} {128}},\
  \bibinfo {pages} {40001} (\bibinfo {year} {2020})}\BibitemShut {NoStop}%
\bibitem [{\citenamefont {Kowalik}\ \emph {et~al.}(2019)\citenamefont
  {Kowalik}, \citenamefont {Daldrop}, \citenamefont {Kappler}, \citenamefont
  {Schulz}, \citenamefont {Schlaich},\ and\ \citenamefont
  {Netz}}]{Kowalik:PRE2019}%
  \BibitemOpen
  \bibfield  {author} {\bibinfo {author} {\bibfnamefont {B.}~\bibnamefont
  {Kowalik}}, \bibinfo {author} {\bibfnamefont {J.~O.}\ \bibnamefont
  {Daldrop}}, \bibinfo {author} {\bibfnamefont {J.}~\bibnamefont {Kappler}},
  \bibinfo {author} {\bibfnamefont {J.~C.~F.}\ \bibnamefont {Schulz}}, \bibinfo
  {author} {\bibfnamefont {A.}~\bibnamefont {Schlaich}},\ and\ \bibinfo
  {author} {\bibfnamefont {R.~R.}\ \bibnamefont {Netz}},\ }\bibfield  {title}
  {\enquote {\bibinfo {title} {Memory-kernel extraction for different molecular
  solutes in solvents of varying viscosity in confinement},}\ }\href
  {https://doi.org/10.1103/PhysRevE.100.012126} {\bibfield  {journal} {\bibinfo
   {journal} {Phys. Rev. E}\ }\textbf {\bibinfo {volume} {100}},\ \bibinfo
  {pages} {012126} (\bibinfo {year} {2019})}\BibitemShut {NoStop}%
\bibitem [{\citenamefont {Straube}\ \emph {et~al.}(2020)\citenamefont
  {Straube}, \citenamefont {Kowalik}, \citenamefont {Netz},\ and\ \citenamefont
  {Höfling}}]{Straube:CP2020}%
  \BibitemOpen
  \bibfield  {author} {\bibinfo {author} {\bibfnamefont {A.~V.}\ \bibnamefont
  {Straube}}, \bibinfo {author} {\bibfnamefont {B.~G.}\ \bibnamefont
  {Kowalik}}, \bibinfo {author} {\bibfnamefont {R.~R.}\ \bibnamefont {Netz}},\
  and\ \bibinfo {author} {\bibfnamefont {F.}~\bibnamefont {Höfling}},\
  }\bibfield  {title} {\enquote {\bibinfo {title} {Rapid onset of molecular
  friction in liquids bridging between the atomistic and hydrodynamic
  pictures},}\ }\href {https://doi.org/10.1038/s42005-020-0389-0} {\bibfield
  {journal} {\bibinfo  {journal} {Commun. Phys.}\ }\textbf {\bibinfo {volume}
  {3}},\ \bibinfo {pages} {126} (\bibinfo {year} {2020})}\BibitemShut {NoStop}%
\bibitem [{\citenamefont {Tassieri}\ \emph {et~al.}(2012)\citenamefont
  {Tassieri}, \citenamefont {Evans}, \citenamefont {Warren}, \citenamefont
  {Bailey},\ and\ \citenamefont {Cooper}}]{Tassieri:NJP2012}%
  \BibitemOpen
  \bibfield  {author} {\bibinfo {author} {\bibfnamefont {M.}~\bibnamefont
  {Tassieri}}, \bibinfo {author} {\bibfnamefont {R.~M.~L.}\ \bibnamefont
  {Evans}}, \bibinfo {author} {\bibfnamefont {R.~L.}\ \bibnamefont {Warren}},
  \bibinfo {author} {\bibfnamefont {N.~J.}\ \bibnamefont {Bailey}},\ and\
  \bibinfo {author} {\bibfnamefont {J.~M.}\ \bibnamefont {Cooper}},\ }\bibfield
   {title} {\enquote {\bibinfo {title} {Microrheology with optical tweezers:
  data analysis},}\ }\href {https://doi.org/10.1088/1367-2630/14/11/115032}
  {\bibfield  {journal} {\bibinfo  {journal} {New J. Phys.}\ }\textbf {\bibinfo
  {volume} {14}},\ \bibinfo {pages} {115032} (\bibinfo {year}
  {2012})}\BibitemShut {NoStop}%
\bibitem [{\citenamefont {Rivas-Barbosa}\ \emph {et~al.}(2020)\citenamefont
  {Rivas-Barbosa}, \citenamefont {Escobedo-Sánchez}, \citenamefont
  {Tassieri},\ and\ \citenamefont {Laurati}}]{RivasBarbosa:PCCP2020}%
  \BibitemOpen
  \bibfield  {author} {\bibinfo {author} {\bibfnamefont {R.}~\bibnamefont
  {Rivas-Barbosa}}, \bibinfo {author} {\bibfnamefont {M.~A.}\ \bibnamefont
  {Escobedo-Sánchez}}, \bibinfo {author} {\bibfnamefont {M.}~\bibnamefont
  {Tassieri}},\ and\ \bibinfo {author} {\bibfnamefont {M.}~\bibnamefont
  {Laurati}},\ }\bibfield  {title} {\enquote {\bibinfo {title} {{i-Rheo}:
  determining the linear viscoelastic moduli of colloidal dispersions from
  step-stress measurements},}\ }\href {https://doi.org/10.1039/c9cp06191f}
  {\bibfield  {journal} {\bibinfo  {journal} {Phys. Chem. Chem. Phys.}\
  }\textbf {\bibinfo {volume} {22}},\ \bibinfo {pages} {3839} (\bibinfo {year}
  {2020})}\BibitemShut {NoStop}%
\bibitem [{\citenamefont {Nishi}\ \emph {et~al.}(2018)\citenamefont {Nishi},
  \citenamefont {Kilfoil}, \citenamefont {Schmidt},\ and\ \citenamefont
  {MacKintosh}}]{Nishi:SM2018}%
  \BibitemOpen
  \bibfield  {author} {\bibinfo {author} {\bibfnamefont {K.}~\bibnamefont
  {Nishi}}, \bibinfo {author} {\bibfnamefont {M.~L.}\ \bibnamefont {Kilfoil}},
  \bibinfo {author} {\bibfnamefont {C.~F.}\ \bibnamefont {Schmidt}},\ and\
  \bibinfo {author} {\bibfnamefont {F.~C.}\ \bibnamefont {MacKintosh}},\
  }\bibfield  {title} {\enquote {\bibinfo {title} {A symmetrical method to
  obtain shear moduli from microrheology},}\ }\href
  {https://doi.org/10.1039/c7sm02499a} {\bibfield  {journal} {\bibinfo
  {journal} {Soft Matter}\ }\textbf {\bibinfo {volume} {14}},\ \bibinfo {pages}
  {3716} (\bibinfo {year} {2018})}\BibitemShut {NoStop}%
\bibitem [{\citenamefont {Vroylandt}\ \emph {et~al.}(2022)\citenamefont
  {Vroylandt}, \citenamefont {Goudenège}, \citenamefont {Monmarché},
  \citenamefont {Pietrucci},\ and\ \citenamefont
  {Rotenberg}}]{Vroylandt:PNAS2022}%
  \BibitemOpen
  \bibfield  {author} {\bibinfo {author} {\bibfnamefont {H.}~\bibnamefont
  {Vroylandt}}, \bibinfo {author} {\bibfnamefont {L.}~\bibnamefont
  {Goudenège}}, \bibinfo {author} {\bibfnamefont {P.}~\bibnamefont
  {Monmarché}}, \bibinfo {author} {\bibfnamefont {F.}~\bibnamefont
  {Pietrucci}},\ and\ \bibinfo {author} {\bibfnamefont {B.}~\bibnamefont
  {Rotenberg}},\ }\bibfield  {title} {\enquote {\bibinfo {title}
  {Likelihood-based non-{M}arkovian models from molecular dynamics},}\ }\href
  {https://doi.org/10.1073/pnas.2117586119} {\bibfield  {journal} {\bibinfo
  {journal} {Proc. Natl. Acad. Sci.}\ }\textbf {\bibinfo {volume} {119}},\
  \bibinfo {pages} {e2117586119} (\bibinfo {year} {2022})}\BibitemShut
  {NoStop}%
\bibitem [{\citenamefont {Lapolla}\ and\ \citenamefont
  {Godec}(2021)}]{Lapolla:PRR2021}%
  \BibitemOpen
  \bibfield  {author} {\bibinfo {author} {\bibfnamefont {A.}~\bibnamefont
  {Lapolla}}\ and\ \bibinfo {author} {\bibfnamefont {A.}~\bibnamefont
  {Godec}},\ }\bibfield  {title} {\enquote {\bibinfo {title} {Toolbox for
  quantifying memory in dynamics along reaction coordinates},}\ }\href
  {https://doi.org/10.1103/physrevresearch.3.l022018} {\bibfield  {journal}
  {\bibinfo  {journal} {Phys. Rev. Res.}\ }\textbf {\bibinfo {volume} {3}},\
  \bibinfo {pages} {L022018} (\bibinfo {year} {2021})}\BibitemShut {NoStop}%
\bibitem [{\citenamefont {Evstigneev}\ \emph {et~al.}(2008)\citenamefont
  {Evstigneev}, \citenamefont {Zvyagolskaya}, \citenamefont {Bleil},
  \citenamefont {Eichhorn}, \citenamefont {Bechinger},\ and\ \citenamefont
  {Reimann}}]{Evstigneev-etal:PRE2008}%
  \BibitemOpen
  \bibfield  {author} {\bibinfo {author} {\bibfnamefont {M.}~\bibnamefont
  {Evstigneev}}, \bibinfo {author} {\bibfnamefont {O.}~\bibnamefont
  {Zvyagolskaya}}, \bibinfo {author} {\bibfnamefont {S.}~\bibnamefont {Bleil}},
  \bibinfo {author} {\bibfnamefont {R.}~\bibnamefont {Eichhorn}}, \bibinfo
  {author} {\bibfnamefont {C.}~\bibnamefont {Bechinger}},\ and\ \bibinfo
  {author} {\bibfnamefont {P.}~\bibnamefont {Reimann}},\ }\bibfield  {title}
  {\enquote {\bibinfo {title} {Diffusion of colloidal particles in a tilted
  periodic potential: Theory versus experiment},}\ }\href
  {https://doi.org/10.1103/PhysRevE.77.041107} {\bibfield  {journal} {\bibinfo
  {journal} {Phys. Rev. E}\ }\textbf {\bibinfo {volume} {77}},\ \bibinfo
  {pages} {041107} (\bibinfo {year} {2008})}\BibitemShut {NoStop}%
\bibitem [{\citenamefont {Straube}\ and\ \citenamefont
  {Tierno}(2013)}]{Straube-Tierno:EPL2013}%
  \BibitemOpen
  \bibfield  {author} {\bibinfo {author} {\bibfnamefont {A.~V.}\ \bibnamefont
  {Straube}}\ and\ \bibinfo {author} {\bibfnamefont {P.}~\bibnamefont
  {Tierno}},\ }\bibfield  {title} {\enquote {\bibinfo {title} {Synchronous vs.
  asynchronous transport of a paramagnetic particle in a modulated ratchet
  potential},}\ }\href {https://doi.org/10.1209/0295-5075/103/28001} {\bibfield
   {journal} {\bibinfo  {journal} {{EPL} (Europhys. Lett.)}\ }\textbf {\bibinfo
  {volume} {103}},\ \bibinfo {pages} {28001} (\bibinfo {year}
  {2013})}\BibitemShut {NoStop}%
\bibitem [{\citenamefont {Straube}\ and\ \citenamefont
  {Tierno}(2014)}]{Straube:SM2014}%
  \BibitemOpen
  \bibfield  {author} {\bibinfo {author} {\bibfnamefont {A.~V.}\ \bibnamefont
  {Straube}}\ and\ \bibinfo {author} {\bibfnamefont {P.}~\bibnamefont
  {Tierno}},\ }\bibfield  {title} {\enquote {\bibinfo {title} {Tunable
  interactions between paramagnetic colloidal particles driven in a modulated
  ratchet potential},}\ }\href {https://doi.org/10.1039/C4SM00132J} {\bibfield
  {journal} {\bibinfo  {journal} {Soft Matter}\ }\textbf {\bibinfo {volume}
  {10}},\ \bibinfo {pages} {3915} (\bibinfo {year} {2014})}\BibitemShut
  {NoStop}%
\bibitem [{\citenamefont {Juniper}\ \emph {et~al.}(2016)\citenamefont
  {Juniper}, \citenamefont {Straube}, \citenamefont {Aarts},\ and\
  \citenamefont {Dullens}}]{Juniper-etal:PRE2016}%
  \BibitemOpen
  \bibfield  {author} {\bibinfo {author} {\bibfnamefont {M.~P.~N.}\
  \bibnamefont {Juniper}}, \bibinfo {author} {\bibfnamefont {A.~V.}\
  \bibnamefont {Straube}}, \bibinfo {author} {\bibfnamefont {D.~G. A.~L.}\
  \bibnamefont {Aarts}},\ and\ \bibinfo {author} {\bibfnamefont {R.~P.~A.}\
  \bibnamefont {Dullens}},\ }\bibfield  {title} {\enquote {\bibinfo {title}
  {Colloidal particles driven across periodic optical-potential-energy
  landscapes},}\ }\href {https://doi.org/10.1103/physreve.93.012608} {\bibfield
   {journal} {\bibinfo  {journal} {Phys. Rev. E}\ }\textbf {\bibinfo {volume}
  {93}},\ \bibinfo {pages} {012608} (\bibinfo {year} {2016})}\BibitemShut
  {NoStop}%
\bibitem [{\citenamefont {Juniper}\ \emph {et~al.}(2017)\citenamefont
  {Juniper}, \citenamefont {Zimmermann}, \citenamefont {Straube}, \citenamefont
  {Besseling}, \citenamefont {Aarts}, \citenamefont {Löwen},\ and\
  \citenamefont {Dullens}}]{Juniper:NJP2017}%
  \BibitemOpen
  \bibfield  {author} {\bibinfo {author} {\bibfnamefont {M.~P.~N.}\
  \bibnamefont {Juniper}}, \bibinfo {author} {\bibfnamefont {U.}~\bibnamefont
  {Zimmermann}}, \bibinfo {author} {\bibfnamefont {A.~V.}\ \bibnamefont
  {Straube}}, \bibinfo {author} {\bibfnamefont {R.}~\bibnamefont {Besseling}},
  \bibinfo {author} {\bibfnamefont {D.~G. A.~L.}\ \bibnamefont {Aarts}},
  \bibinfo {author} {\bibfnamefont {H.}~\bibnamefont {Löwen}},\ and\ \bibinfo
  {author} {\bibfnamefont {R.~P.~A.}\ \bibnamefont {Dullens}},\ }\bibfield
  {title} {\enquote {\bibinfo {title} {Dynamic mode locking in a driven
  colloidal system: experiments and theory},}\ }\href
  {https://doi.org/10.1088/1367-2630/aa53cd} {\bibfield  {journal} {\bibinfo
  {journal} {New J. Phys.}\ }\textbf {\bibinfo {volume} {19}},\ \bibinfo
  {pages} {013010} (\bibinfo {year} {2017})}\BibitemShut {NoStop}%
\bibitem [{\citenamefont {Stoop}, \citenamefont {Straube},\ and\ \citenamefont
  {Tierno}(2018)}]{Stoop-etal:NL2018}%
  \BibitemOpen
  \bibfield  {author} {\bibinfo {author} {\bibfnamefont {R.~L.}\ \bibnamefont
  {Stoop}}, \bibinfo {author} {\bibfnamefont {A.~V.}\ \bibnamefont {Straube}},\
  and\ \bibinfo {author} {\bibfnamefont {P.}~\bibnamefont {Tierno}},\
  }\bibfield  {title} {\enquote {\bibinfo {title} {Enhancing nanoparticle
  diffusion on a unidirectional domain wall magnetic ratchet},}\ }\href
  {https://doi.org/10.1021/acs.nanolett.8b04248} {\bibfield  {journal}
  {\bibinfo  {journal} {Nano Lett.}\ }\textbf {\bibinfo {volume} {19}},\
  \bibinfo {pages} {433} (\bibinfo {year} {2018})}\BibitemShut {NoStop}%
\bibitem [{\citenamefont {Stoop}\ \emph {et~al.}(2020)\citenamefont {Stoop},
  \citenamefont {Straube}, \citenamefont {Johansen},\ and\ \citenamefont
  {Tierno}}]{Stoop-etal:PRL2020}%
  \BibitemOpen
  \bibfield  {author} {\bibinfo {author} {\bibfnamefont {R.~L.}\ \bibnamefont
  {Stoop}}, \bibinfo {author} {\bibfnamefont {A.~V.}\ \bibnamefont {Straube}},
  \bibinfo {author} {\bibfnamefont {T.~H.}\ \bibnamefont {Johansen}},\ and\
  \bibinfo {author} {\bibfnamefont {P.}~\bibnamefont {Tierno}},\ }\bibfield
  {title} {\enquote {\bibinfo {title} {Collective directional locking of
  colloidal monolayers on a periodic substrate},}\ }\href
  {https://doi.org/10.1103/physrevlett.124.058002} {\bibfield  {journal}
  {\bibinfo  {journal} {Phys. Rev. Lett.}\ }\textbf {\bibinfo {volume} {124}},\
  \bibinfo {pages} {058002} (\bibinfo {year} {2020})}\BibitemShut {NoStop}%
\bibitem [{\citenamefont {Ma}\ \emph {et~al.}(2015)\citenamefont {Ma},
  \citenamefont {Lai}, \citenamefont {Ackerson},\ and\ \citenamefont
  {Tong}}]{Ma:SM2015}%
  \BibitemOpen
  \bibfield  {author} {\bibinfo {author} {\bibfnamefont {X.-g.}\ \bibnamefont
  {Ma}}, \bibinfo {author} {\bibfnamefont {P.-Y.}\ \bibnamefont {Lai}},
  \bibinfo {author} {\bibfnamefont {B.~J.}\ \bibnamefont {Ackerson}},\ and\
  \bibinfo {author} {\bibfnamefont {P.}~\bibnamefont {Tong}},\ }\bibfield
  {title} {\enquote {\bibinfo {title} {Colloidal transport and diffusion over a
  tilted periodic potential: dynamics of individual particles},}\ }\href
  {https://doi.org/10.1039/C4SM02387K} {\bibfield  {journal} {\bibinfo
  {journal} {Soft Matter}\ }\textbf {\bibinfo {volume} {11}},\ \bibinfo {pages}
  {1182} (\bibinfo {year} {2015})}\BibitemShut {NoStop}%
\bibitem [{\citenamefont {Su}\ \emph {et~al.}(2017)\citenamefont {Su},
  \citenamefont {Lai}, \citenamefont {Ackerson}, \citenamefont {Cao},
  \citenamefont {Han},\ and\ \citenamefont {Tong}}]{Su:JCP2017}%
  \BibitemOpen
  \bibfield  {author} {\bibinfo {author} {\bibfnamefont {Y.}~\bibnamefont
  {Su}}, \bibinfo {author} {\bibfnamefont {P.-Y.}\ \bibnamefont {Lai}},
  \bibinfo {author} {\bibfnamefont {B.~J.}\ \bibnamefont {Ackerson}}, \bibinfo
  {author} {\bibfnamefont {X.}~\bibnamefont {Cao}}, \bibinfo {author}
  {\bibfnamefont {Y.}~\bibnamefont {Han}},\ and\ \bibinfo {author}
  {\bibfnamefont {P.}~\bibnamefont {Tong}},\ }\bibfield  {title} {\enquote
  {\bibinfo {title} {Colloidal diffusion over a quasicrystalline-patterned
  surface},}\ }\href {https://doi.org/10.1063/1.4984938} {\bibfield  {journal}
  {\bibinfo  {journal} {J. Chem. Phys.}\ }\textbf {\bibinfo {volume} {146}},\
  \bibinfo {pages} {214903} (\bibinfo {year} {2017})}\BibitemShut {NoStop}%
\bibitem [{\citenamefont {Choudhury}\ \emph {et~al.}(2017)\citenamefont
  {Choudhury}, \citenamefont {Straube}, \citenamefont {Fischer}, \citenamefont
  {Gibbs},\ and\ \citenamefont {Höfling}}]{Choudhury:NJP2017}%
  \BibitemOpen
  \bibfield  {author} {\bibinfo {author} {\bibfnamefont {U.}~\bibnamefont
  {Choudhury}}, \bibinfo {author} {\bibfnamefont {A.~V.}\ \bibnamefont
  {Straube}}, \bibinfo {author} {\bibfnamefont {P.}~\bibnamefont {Fischer}},
  \bibinfo {author} {\bibfnamefont {J.~G.}\ \bibnamefont {Gibbs}},\ and\
  \bibinfo {author} {\bibfnamefont {F.}~\bibnamefont {Höfling}},\ }\bibfield
  {title} {\enquote {\bibinfo {title} {Active colloidal propulsion over a
  crystalline surface},}\ }\href {https://doi.org/10.1088/1367-2630/aa9b4b}
  {\bibfield  {journal} {\bibinfo  {journal} {New J. Phys.}\ }\textbf {\bibinfo
  {volume} {19}},\ \bibinfo {pages} {125010} (\bibinfo {year}
  {2017})}\BibitemShut {NoStop}%
\bibitem [{\citenamefont {Berner}\ \emph {et~al.}(2018)\citenamefont {Berner},
  \citenamefont {Müller}, \citenamefont {Gomez-Solano}, \citenamefont
  {Krüger},\ and\ \citenamefont {Bechinger}}]{Berner:NC2018}%
  \BibitemOpen
  \bibfield  {author} {\bibinfo {author} {\bibfnamefont {J.}~\bibnamefont
  {Berner}}, \bibinfo {author} {\bibfnamefont {B.}~\bibnamefont {Müller}},
  \bibinfo {author} {\bibfnamefont {J.~R.}\ \bibnamefont {Gomez-Solano}},
  \bibinfo {author} {\bibfnamefont {M.}~\bibnamefont {Krüger}},\ and\ \bibinfo
  {author} {\bibfnamefont {C.}~\bibnamefont {Bechinger}},\ }\bibfield  {title}
  {\enquote {\bibinfo {title} {Oscillating modes of driven colloids in
  overdamped systems},}\ }\href {https://doi.org/10.1038/s41467-018-03345-2}
  {\bibfield  {journal} {\bibinfo  {journal} {Nat. Commun.}\ }\textbf {\bibinfo
  {volume} {9}},\ \bibinfo {pages} {999} (\bibinfo {year} {2018})}\BibitemShut
  {NoStop}%
\bibitem [{\citenamefont {Reimann}\ \emph {et~al.}(2002)\citenamefont
  {Reimann}, \citenamefont {Van~den Broeck}, \citenamefont {Linke},
  \citenamefont {Hänggi}, \citenamefont {Rubi},\ and\ \citenamefont
  {Pérez-Madrid}}]{Reimann:PRE2002}%
  \BibitemOpen
  \bibfield  {author} {\bibinfo {author} {\bibfnamefont {P.}~\bibnamefont
  {Reimann}}, \bibinfo {author} {\bibfnamefont {C.}~\bibnamefont {Van~den
  Broeck}}, \bibinfo {author} {\bibfnamefont {H.}~\bibnamefont {Linke}},
  \bibinfo {author} {\bibfnamefont {P.}~\bibnamefont {Hänggi}}, \bibinfo
  {author} {\bibfnamefont {J.~M.}\ \bibnamefont {Rubi}},\ and\ \bibinfo
  {author} {\bibfnamefont {A.}~\bibnamefont {Pérez-Madrid}},\ }\bibfield
  {title} {\enquote {\bibinfo {title} {Diffusion in tilted periodic potentials:
  {E}nhancement, universality, and scaling},}\ }\href
  {https://doi.org/10.1103/physreve.65.031104} {\bibfield  {journal} {\bibinfo
  {journal} {Phys. Rev. E}\ }\textbf {\bibinfo {volume} {65}},\ \bibinfo
  {pages} {031104} (\bibinfo {year} {2002})}\BibitemShut {NoStop}%
\bibitem [{\citenamefont {Lapolla}\ and\ \citenamefont
  {Godec}(2019)}]{Lapolla:FP2019}%
  \BibitemOpen
  \bibfield  {author} {\bibinfo {author} {\bibfnamefont {A.}~\bibnamefont
  {Lapolla}}\ and\ \bibinfo {author} {\bibfnamefont {A.}~\bibnamefont
  {Godec}},\ }\bibfield  {title} {\enquote {\bibinfo {title} {Manifestations of
  projection-induced memory: General theory and the tilted single file},}\
  }\href {https://doi.org/10.3389/fphy.2019.00182} {\bibfield  {journal}
  {\bibinfo  {journal} {Front. Phys.}\ }\textbf {\bibinfo {volume} {7}},\
  \bibinfo {pages} {182} (\bibinfo {year} {2019})}\BibitemShut {NoStop}%
\bibitem [{\citenamefont {Leitmann}, \citenamefont {Bénichou},\ and\
  \citenamefont {Franosch}(2018)}]{LeitmannJPAMT2018}%
  \BibitemOpen
  \bibfield  {author} {\bibinfo {author} {\bibfnamefont {S.}~\bibnamefont
  {Leitmann}}, \bibinfo {author} {\bibfnamefont {O.}~\bibnamefont
  {Bénichou}},\ and\ \bibinfo {author} {\bibfnamefont {T.}~\bibnamefont
  {Franosch}},\ }\bibfield  {title} {\enquote {\bibinfo {title} {Time-dependent
  dynamics of the three-dimensional driven lattice {L}orentz gas},}\ }\href
  {https://doi.org/10.1088/1751-8121/aad341} {\bibfield  {journal} {\bibinfo
  {journal} {J. Phys. A: Math. Theor.}\ }\textbf {\bibinfo {volume} {51}},\
  \bibinfo {pages} {375001} (\bibinfo {year} {2018})}\BibitemShut {NoStop}%
\bibitem [{\citenamefont {Leitmann}\ and\ \citenamefont
  {Franosch}(2017)}]{LeitmannPRL2017}%
  \BibitemOpen
  \bibfield  {author} {\bibinfo {author} {\bibfnamefont {S.}~\bibnamefont
  {Leitmann}}\ and\ \bibinfo {author} {\bibfnamefont {T.}~\bibnamefont
  {Franosch}},\ }\bibfield  {title} {\enquote {\bibinfo {title} {Time-dependent
  fluctuations and superdiffusivity in the driven lattice {L}orentz gas},}\
  }\href {https://doi.org/10.1103/PhysRevLett.118.018001} {\bibfield  {journal}
  {\bibinfo  {journal} {Phys. Rev. Lett.}\ }\textbf {\bibinfo {volume} {118}},\
  \bibinfo {pages} {018001} (\bibinfo {year} {2017})}\BibitemShut {NoStop}%
\bibitem [{\citenamefont {Leitmann}\ and\ \citenamefont
  {Franosch}(2013)}]{LeitmannPRL2013}%
  \BibitemOpen
  \bibfield  {author} {\bibinfo {author} {\bibfnamefont {S.}~\bibnamefont
  {Leitmann}}\ and\ \bibinfo {author} {\bibfnamefont {T.}~\bibnamefont
  {Franosch}},\ }\bibfield  {title} {\enquote {\bibinfo {title} {Nonlinear
  response in the driven lattice {L}orentz gas},}\ }\href
  {https://doi.org/10.1103/PhysRevLett.111.190603} {\bibfield  {journal}
  {\bibinfo  {journal} {Phys. Rev. Lett.}\ }\textbf {\bibinfo {volume} {111}},\
  \bibinfo {pages} {190603} (\bibinfo {year} {2013})}\BibitemShut {NoStop}%
\bibitem [{\citenamefont {Lifson}\ and\ \citenamefont
  {Jackson}(1962)}]{Lifson:JCP1962}%
  \BibitemOpen
  \bibfield  {author} {\bibinfo {author} {\bibfnamefont {S.}~\bibnamefont
  {Lifson}}\ and\ \bibinfo {author} {\bibfnamefont {J.~L.}\ \bibnamefont
  {Jackson}},\ }\bibfield  {title} {\enquote {\bibinfo {title} {On the
  self-diffusion of ions in a polyelectrolyte solution},}\ }\href
  {https://doi.org/10.1063/1.1732899} {\bibfield  {journal} {\bibinfo
  {journal} {J. Chem. Phys.}\ }\textbf {\bibinfo {volume} {36}},\ \bibinfo
  {pages} {2410} (\bibinfo {year} {1962})}\BibitemShut {NoStop}%
\bibitem [{\citenamefont {Festa}\ and\ \citenamefont
  {d’Agliano}(1978)}]{Festa:PA1978}%
  \BibitemOpen
  \bibfield  {author} {\bibinfo {author} {\bibfnamefont {R.}~\bibnamefont
  {Festa}}\ and\ \bibinfo {author} {\bibfnamefont {E.}~\bibnamefont
  {d’Agliano}},\ }\bibfield  {title} {\enquote {\bibinfo {title} {Diffusion
  coefficient for a {B}rownian particle in a periodic field of force},}\ }\href
  {https://doi.org/10.1016/0378-4371(78)90111-5} {\bibfield  {journal}
  {\bibinfo  {journal} {Physica A: Statistical Mechanics and its Applications}\
  }\textbf {\bibinfo {volume} {90}},\ \bibinfo {pages} {229} (\bibinfo {year}
  {1978})}\BibitemShut {NoStop}%
\bibitem [{\citenamefont {Stratonovich}(1967)}]{Stratonovich:1967}%
  \BibitemOpen
  \bibfield  {author} {\bibinfo {author} {\bibfnamefont {R.~L.}\ \bibnamefont
  {Stratonovich}},\ }\href@noop {} {\emph {\bibinfo {title} {Topics in the
  Theory of Random Noise}}},\ Vol.~\bibinfo {volume} {II}\ (\bibinfo
  {publisher} {Gordon and Breach},\ \bibinfo {address} {New York},\ \bibinfo
  {year} {1967})\BibitemShut {NoStop}%
\bibitem [{\citenamefont {Reimann}\ \emph {et~al.}(2001)\citenamefont
  {Reimann}, \citenamefont {Van~den Broeck}, \citenamefont {Linke},
  \citenamefont {Hänggi}, \citenamefont {Rubi},\ and\ \citenamefont
  {Pérez-Madrid}}]{Reimann:PRL2001}%
  \BibitemOpen
  \bibfield  {author} {\bibinfo {author} {\bibfnamefont {P.}~\bibnamefont
  {Reimann}}, \bibinfo {author} {\bibfnamefont {C.}~\bibnamefont {Van~den
  Broeck}}, \bibinfo {author} {\bibfnamefont {H.}~\bibnamefont {Linke}},
  \bibinfo {author} {\bibfnamefont {P.}~\bibnamefont {Hänggi}}, \bibinfo
  {author} {\bibfnamefont {J.~M.}\ \bibnamefont {Rubi}},\ and\ \bibinfo
  {author} {\bibfnamefont {A.}~\bibnamefont {Pérez-Madrid}},\ }\bibfield
  {title} {\enquote {\bibinfo {title} {Giant acceleration of free diffusion by
  use of tilted periodic potentials},}\ }\href
  {https://doi.org/10.1103/physrevlett.87.010602} {\bibfield  {journal}
  {\bibinfo  {journal} {Phys. Rev. Lett.}\ }\textbf {\bibinfo {volume} {87}},\
  \bibinfo {pages} {010602} (\bibinfo {year} {2001})}\BibitemShut {NoStop}%
\bibitem [{\citenamefont {Post}, \citenamefont {Wolf},\ and\ \citenamefont
  {Stock}(2022)}]{Post:JCTC2022}%
  \BibitemOpen
  \bibfield  {author} {\bibinfo {author} {\bibfnamefont {M.}~\bibnamefont
  {Post}}, \bibinfo {author} {\bibfnamefont {S.}~\bibnamefont {Wolf}},\ and\
  \bibinfo {author} {\bibfnamefont {G.}~\bibnamefont {Stock}},\ }\bibfield
  {title} {\enquote {\bibinfo {title} {Molecular origin of driving-dependent
  friction in fluids},}\ }\href {https://doi.org/10.1021/acs.jctc.2c00190}
  {\bibfield  {journal} {\bibinfo  {journal} {J. Chem. Theory Comput.}\
  }\textbf {\bibinfo {volume} {18}},\ \bibinfo {pages} {2816} (\bibinfo {year}
  {2022})}\BibitemShut {NoStop}%
\bibitem [{\citenamefont {Kubo}, \citenamefont {Toda},\ and\ \citenamefont
  {Hashitsume}(1991)}]{Kubo:Book1991}%
  \BibitemOpen
  \bibfield  {author} {\bibinfo {author} {\bibfnamefont {R.}~\bibnamefont
  {Kubo}}, \bibinfo {author} {\bibfnamefont {M.}~\bibnamefont {Toda}},\ and\
  \bibinfo {author} {\bibfnamefont {N.}~\bibnamefont {Hashitsume}},\
  }\href@noop {} {\emph {\bibinfo {title} {Statistical Physics~{II:}
  {Nonequilibrium} Statistical Mechanics}}}\ (\bibinfo  {publisher}
  {Springer},\ \bibinfo {address} {Berlin, Heidelberg},\ \bibinfo {year}
  {1991})\BibitemShut {NoStop}%
\bibitem [{\citenamefont {Teschl}(2014)}]{Teschl:MathMeth}%
  \BibitemOpen
  \bibfield  {author} {\bibinfo {author} {\bibfnamefont {G.}~\bibnamefont
  {Teschl}},\ }\href@noop {} {\emph {\bibinfo {title} {Mathematical methods in
  quantum mechanics: with applications to {Schrödinger} operators}}},\
  \bibinfo {edition} {2nd}\ ed.\ (\bibinfo  {publisher} {Am. Math. Soc.},\
  \bibinfo {address} {Providence, RI},\ \bibinfo {year} {2014})\BibitemShut
  {NoStop}%
\bibitem [{\citenamefont {Franosch}(2014)}]{Franosch:JPA2014}%
  \BibitemOpen
  \bibfield  {author} {\bibinfo {author} {\bibfnamefont {T.}~\bibnamefont
  {Franosch}},\ }\bibfield  {title} {\enquote {\bibinfo {title} {Long-time
  limit of correlation functions},}\ }\href
  {http://stacks.iop.org/1751-8121/47/i=32/a=325004} {\bibfield  {journal}
  {\bibinfo  {journal} {J. Phys. A: Math. Theor.}\ }\textbf {\bibinfo {volume}
  {47}},\ \bibinfo {pages} {325004} (\bibinfo {year} {2014})}\BibitemShut
  {NoStop}%
\bibitem [{\citenamefont {Bauer}\ \emph {et~al.}(2010)\citenamefont {Bauer},
  \citenamefont {H{\"o}fling}, \citenamefont {Munk}, \citenamefont {Frey},\
  and\ \citenamefont {Franosch}}]{Bauer:EPJST2010}%
  \BibitemOpen
  \bibfield  {author} {\bibinfo {author} {\bibfnamefont {T.}~\bibnamefont
  {Bauer}}, \bibinfo {author} {\bibfnamefont {F.}~\bibnamefont {H{\"o}fling}},
  \bibinfo {author} {\bibfnamefont {T.}~\bibnamefont {Munk}}, \bibinfo {author}
  {\bibfnamefont {E.}~\bibnamefont {Frey}},\ and\ \bibinfo {author}
  {\bibfnamefont {T.}~\bibnamefont {Franosch}},\ }\bibfield  {title} {\enquote
  {\bibinfo {title} {The localization transition of the two-dimensional
  {L}orentz model},}\ }\href {https://doi.org/10.1140/epjst/e2010-01313-1}
  {\bibfield  {journal} {\bibinfo  {journal} {Eur. Phys. J. Special Topics}\
  }\textbf {\bibinfo {volume} {189}},\ \bibinfo {pages} {103} (\bibinfo {year}
  {2010})}\BibitemShut {NoStop}%
\bibitem [{\citenamefont {Panja}(2010{\natexlab{a}})}]{Panja:JSM2010}%
  \BibitemOpen
  \bibfield  {author} {\bibinfo {author} {\bibfnamefont {D.}~\bibnamefont
  {Panja}},\ }\bibfield  {title} {\enquote {\bibinfo {title} {Generalized
  {L}angevin equation formulation for anomalous polymer dynamics},}\ }\href
  {https://doi.org/10.1088/1742-5468/2010/02/l02001} {\bibfield  {journal}
  {\bibinfo  {journal} {J. Stat. Mech.: Theory Exp.}\ }\textbf {\bibinfo
  {volume} {2010}},\ \bibinfo {pages} {L02001} (\bibinfo {year}
  {2010}{\natexlab{a}})}\BibitemShut {NoStop}%
\bibitem [{\citenamefont {Panja}(2010{\natexlab{b}})}]{Panja:JSM2010a}%
  \BibitemOpen
  \bibfield  {author} {\bibinfo {author} {\bibfnamefont {D.}~\bibnamefont
  {Panja}},\ }\bibfield  {title} {\enquote {\bibinfo {title} {Anomalous polymer
  dynamics is non-{M}arkovian: memory effects and the generalized {L}angevin
  equation formulation},}\ }\href
  {https://doi.org/10.1088/1742-5468/2010/06/p06011} {\bibfield  {journal}
  {\bibinfo  {journal} {J. Stat. Mech.: Theory Exp.}\ }\textbf {\bibinfo
  {volume} {2010}},\ \bibinfo {pages} {P06011} (\bibinfo {year}
  {2010}{\natexlab{b}})}\BibitemShut {NoStop}%
\bibitem [{\citenamefont {Miyaguchi}(2022)}]{Miyaguchi:PRR2022}%
  \BibitemOpen
  \bibfield  {author} {\bibinfo {author} {\bibfnamefont {T.}~\bibnamefont
  {Miyaguchi}},\ }\bibfield  {title} {\enquote {\bibinfo {title} {Generalized
  {L}angevin equation with fluctuating diffusivity},}\ }\href
  {https://doi.org/10.1103/physrevresearch.4.043062} {\bibfield  {journal}
  {\bibinfo  {journal} {Phys. Rev. Research}\ }\textbf {\bibinfo {volume}
  {4}},\ \bibinfo {pages} {043062} (\bibinfo {year} {2022})}\BibitemShut
  {NoStop}%
\bibitem [{\citenamefont {Jones}(1961)}]{Jones:MC1961}%
  \BibitemOpen
  \bibfield  {author} {\bibinfo {author} {\bibfnamefont {J.~G.}\ \bibnamefont
  {Jones}},\ }\bibfield  {title} {\enquote {\bibinfo {title} {On the numerical
  solution of convolution integral equations and systems of such equations},}\
  }\href {https://community.ams.org/journals/mcom/1961-15-074/S0025-
  5718-1961-0122001-7/S0025-5718-1961-0122001-7.pdf} {\bibfield  {journal}
  {\bibinfo  {journal} {Math. Comput.}\ }\textbf {\bibinfo {volume} {15}},\
  \bibinfo {pages} {131} (\bibinfo {year} {1961})}\BibitemShut {NoStop}%
\bibitem [{\citenamefont {Frenkel}\ and\ \citenamefont
  {Smit}(2001)}]{Frenkel:MD}%
  \BibitemOpen
  \bibfield  {author} {\bibinfo {author} {\bibfnamefont {D.}~\bibnamefont
  {Frenkel}}\ and\ \bibinfo {author} {\bibfnamefont {B.~J.}\ \bibnamefont
  {Smit}},\ }\href@noop {} {\emph {\bibinfo {title} {Understanding Molecular
  Simulation}}},\ \bibinfo {edition} {2nd}\ ed.\ (\bibinfo  {publisher}
  {Academic Press},\ \bibinfo {address} {London},\ \bibinfo {year}
  {2001})\BibitemShut {NoStop}%
\bibitem [{\citenamefont {H{\"o}f{}ling}, \citenamefont {Franosch},\ and\
  \citenamefont {Frey}(2006)}]{Hoefling:PRL2006}%
  \BibitemOpen
  \bibfield  {author} {\bibinfo {author} {\bibfnamefont {F.}~\bibnamefont
  {H{\"o}f{}ling}}, \bibinfo {author} {\bibfnamefont {T.}~\bibnamefont
  {Franosch}},\ and\ \bibinfo {author} {\bibfnamefont {E.}~\bibnamefont
  {Frey}},\ }\bibfield  {title} {\enquote {\bibinfo {title} {Localization
  transition of the three-dimensional {L}orentz model and continuum
  percolation},}\ }\href {https://doi.org/10.1103/PhysRevLett.96.165901}
  {\bibfield  {journal} {\bibinfo  {journal} {Phys. Rev. Lett.}\ }\textbf
  {\bibinfo {volume} {96}},\ \bibinfo {pages} {165901} (\bibinfo {year}
  {2006})}\BibitemShut {NoStop}%
\bibitem [{\citenamefont {H{\"o}f{}ling}\ and\ \citenamefont
  {Franosch}(2007)}]{Hoefling:PRL2007}%
  \BibitemOpen
  \bibfield  {author} {\bibinfo {author} {\bibfnamefont {F.}~\bibnamefont
  {H{\"o}f{}ling}}\ and\ \bibinfo {author} {\bibfnamefont {T.}~\bibnamefont
  {Franosch}},\ }\bibfield  {title} {\enquote {\bibinfo {title} {Crossover in
  the slow decay of dynamic correlations in the {L}orentz model},}\ }\href
  {https://doi.org/10.1103/PhysRevLett.98.140601} {\bibfield  {journal}
  {\bibinfo  {journal} {Phys. Rev. Lett.}\ }\textbf {\bibinfo {volume} {98}},\
  \bibinfo {eid} {140601} (\bibinfo {year} {2007})}\BibitemShut {NoStop}%
\bibitem [{\citenamefont {Colberg}\ and\ \citenamefont
  {H{\"o}f{}ling}(2011)}]{Colberg:CPC2011}%
  \BibitemOpen
  \bibfield  {author} {\bibinfo {author} {\bibfnamefont {P.~H.}\ \bibnamefont
  {Colberg}}\ and\ \bibinfo {author} {\bibfnamefont {F.}~\bibnamefont
  {H{\"o}f{}ling}},\ }\bibfield  {title} {\enquote {\bibinfo {title} {Highly
  accelerated simulations of glassy dynamics using {GPU}s: Caveats on limited
  floating-point precision},}\ }\href
  {https://doi.org/10.1016/j.cpc.2011.01.009} {\bibfield  {journal} {\bibinfo
  {journal} {Comput. Phys. Commun.}\ }\textbf {\bibinfo {volume} {182}},\
  \bibinfo {pages} {1120} (\bibinfo {year} {2011})}\BibitemShut {NoStop}%
\bibitem [{\citenamefont {Fulde}\ \emph {et~al.}(1975)\citenamefont {Fulde},
  \citenamefont {Pietronero}, \citenamefont {Schneider},\ and\ \citenamefont
  {Strässler}}]{Fulde:PRL1975}%
  \BibitemOpen
  \bibfield  {author} {\bibinfo {author} {\bibfnamefont {P.}~\bibnamefont
  {Fulde}}, \bibinfo {author} {\bibfnamefont {L.}~\bibnamefont {Pietronero}},
  \bibinfo {author} {\bibfnamefont {W.~R.}\ \bibnamefont {Schneider}},\ and\
  \bibinfo {author} {\bibfnamefont {S.}~\bibnamefont {Strässler}},\ }\bibfield
   {title} {\enquote {\bibinfo {title} {Problem of {B}rownian motion in a
  periodic potential},}\ }\href {https://doi.org/10.1103/PhysRevLett.35.1776}
  {\bibfield  {journal} {\bibinfo  {journal} {Phys. Rev. Lett.}\ }\textbf
  {\bibinfo {volume} {35}},\ \bibinfo {pages} {1776} (\bibinfo {year}
  {1975})}\BibitemShut {NoStop}%
\bibitem [{\citenamefont {Kurzthaler}\ \emph {et~al.}(2018)\citenamefont
  {Kurzthaler}, \citenamefont {Devailly}, \citenamefont {Arlt}, \citenamefont
  {Franosch}, \citenamefont {Poon}, \citenamefont {Martinez},\ and\
  \citenamefont {Brown}}]{Kurzthaler:PRL2018}%
  \BibitemOpen
  \bibfield  {author} {\bibinfo {author} {\bibfnamefont {C.}~\bibnamefont
  {Kurzthaler}}, \bibinfo {author} {\bibfnamefont {C.}~\bibnamefont
  {Devailly}}, \bibinfo {author} {\bibfnamefont {J.}~\bibnamefont {Arlt}},
  \bibinfo {author} {\bibfnamefont {T.}~\bibnamefont {Franosch}}, \bibinfo
  {author} {\bibfnamefont {W.~C.}\ \bibnamefont {Poon}}, \bibinfo {author}
  {\bibfnamefont {V.~A.}\ \bibnamefont {Martinez}},\ and\ \bibinfo {author}
  {\bibfnamefont {A.~T.}\ \bibnamefont {Brown}},\ }\bibfield  {title} {\enquote
  {\bibinfo {title} {Probing the spatiotemporal dynamics of catalytic {J}anus
  particles with single-particle tracking and differential dynamic
  microscopy},}\ }\href {https://doi.org/10.1103/physrevlett.121.078001}
  {\bibfield  {journal} {\bibinfo  {journal} {Phys. Rev. Lett.}\ }\textbf
  {\bibinfo {volume} {121}},\ \bibinfo {pages} {078001} (\bibinfo {year}
  {2018})}\BibitemShut {NoStop}%
\bibitem [{\citenamefont {Kurzthaler}, \citenamefont {Leitmann},\ and\
  \citenamefont {Franosch}(2016)}]{Kurzthaler:SR2016}%
  \BibitemOpen
  \bibfield  {author} {\bibinfo {author} {\bibfnamefont {C.}~\bibnamefont
  {Kurzthaler}}, \bibinfo {author} {\bibfnamefont {S.}~\bibnamefont
  {Leitmann}},\ and\ \bibinfo {author} {\bibfnamefont {T.}~\bibnamefont
  {Franosch}},\ }\bibfield  {title} {\enquote {\bibinfo {title} {Intermediate
  scattering function of an anisotropic active {B}rownian particle},}\ }\href
  {https://doi.org/10.1038/srep36702} {\bibfield  {journal} {\bibinfo
  {journal} {Sci. Rep.}\ }\textbf {\bibinfo {volume} {6}},\ \bibinfo {pages}
  {36702} (\bibinfo {year} {2016})}\BibitemShut {NoStop}%
\bibitem [{\citenamefont {Caraglio}\ and\ \citenamefont
  {Franosch}(2022)}]{Caraglio:PRL2022}%
  \BibitemOpen
  \bibfield  {author} {\bibinfo {author} {\bibfnamefont {M.}~\bibnamefont
  {Caraglio}}\ and\ \bibinfo {author} {\bibfnamefont {T.}~\bibnamefont
  {Franosch}},\ }\bibfield  {title} {\enquote {\bibinfo {title} {Analytic
  solution of an active {B}rownian particle in a harmonic well},}\ }\href
  {https://doi.org/10.1103/physrevlett.129.158001} {\bibfield  {journal}
  {\bibinfo  {journal} {Phys. Rev. Lett.}\ }\textbf {\bibinfo {volume} {129}},\
  \bibinfo {pages} {158001} (\bibinfo {year} {2022})}\BibitemShut {NoStop}%
\bibitem [{\citenamefont {Feller}(1968)}]{Feller:Probability:Vol2}%
  \BibitemOpen
  \bibfield  {author} {\bibinfo {author} {\bibfnamefont {W.}~\bibnamefont
  {Feller}},\ }\href@noop {} {\emph {\bibinfo {title} {An Introduction to
  Probability Theory and Its Applications}}},\ \bibinfo {edition} {3rd}\ ed.,\
  Vol.~\bibinfo {volume} {2}\ (\bibinfo  {publisher} {Wiley},\ \bibinfo {year}
  {1968})\BibitemShut {NoStop}%
\bibitem [{\citenamefont {Franosch}\ and\ \citenamefont
  {Götze}(1999)}]{Franosch:JPCB1999}%
  \BibitemOpen
  \bibfield  {author} {\bibinfo {author} {\bibfnamefont {T.}~\bibnamefont
  {Franosch}}\ and\ \bibinfo {author} {\bibfnamefont {W.}~\bibnamefont
  {Götze}},\ }\bibfield  {title} {\enquote {\bibinfo {title} {Relaxation rate
  distributions for supercooled liquids},}\ }\href
  {https://doi.org/10.1021/jp983412r} {\bibfield  {journal} {\bibinfo
  {journal} {J. Phys. Chem. B}\ }\textbf {\bibinfo {volume} {103}},\ \bibinfo
  {pages} {4011} (\bibinfo {year} {1999})}\BibitemShut {NoStop}%
\bibitem [{\citenamefont {Franosch}\ and\ \citenamefont
  {Voigtmann}(2002)}]{Franosch:JSP2002}%
  \BibitemOpen
  \bibfield  {author} {\bibinfo {author} {\bibfnamefont {T.}~\bibnamefont
  {Franosch}}\ and\ \bibinfo {author} {\bibfnamefont {T.}~\bibnamefont
  {Voigtmann}},\ }\bibfield  {title} {\enquote {\bibinfo {title} {Completely
  monotone solutions of the mode-coupling theory for mixtures},}\ }\href
  {https://doi.org/10.1023/A:1019991729106} {\bibfield  {journal} {\bibinfo
  {journal} {J. Stat. Phys.}\ }\textbf {\bibinfo {volume} {109}},\ \bibinfo
  {pages} {237} (\bibinfo {year} {2002})}\BibitemShut {NoStop}%
\bibitem [{\citenamefont {Hohenegger}\ and\ \citenamefont
  {McKinley}(2018)}]{Hohenegger:SJAM2018}%
  \BibitemOpen
  \bibfield  {author} {\bibinfo {author} {\bibfnamefont {C.}~\bibnamefont
  {Hohenegger}}\ and\ \bibinfo {author} {\bibfnamefont {S.~A.}\ \bibnamefont
  {McKinley}},\ }\bibfield  {title} {\enquote {\bibinfo {title} {Reconstructing
  complex fluid properties from the behavior of fluctuating immersed
  particles},}\ }\href {https://doi.org/10.1137/17m1131660} {\bibfield
  {journal} {\bibinfo  {journal} {SIAM J. Appl. Math.}\ }\textbf {\bibinfo
  {volume} {78}},\ \bibinfo {pages} {2200} (\bibinfo {year}
  {2018})}\BibitemShut {NoStop}%
\bibitem [{\citenamefont {Øksendal}(2010)}]{Oksendal:Stochastic}%
  \BibitemOpen
  \bibfield  {author} {\bibinfo {author} {\bibfnamefont {B.}~\bibnamefont
  {Øksendal}},\ }\href@noop {} {\emph {\bibinfo {title} {Stochastic
  Differential Equations: An Introduction with Applications}}},\ \bibinfo
  {edition} {6th}\ ed.\ (\bibinfo  {publisher} {Springer},\ \bibinfo {address}
  {Berlin},\ \bibinfo {year} {2010})\BibitemShut {NoStop}%
\bibitem [{\citenamefont {Boon}\ and\ \citenamefont
  {Yip}(1991{\natexlab{b}})}]{BoonYip:1980}%
  \BibitemOpen
  \bibfield  {author} {\bibinfo {author} {\bibfnamefont {J.~P.}\ \bibnamefont
  {Boon}}\ and\ \bibinfo {author} {\bibfnamefont {S.}~\bibnamefont {Yip}},\
  }\href@noop {} {\emph {\bibinfo {title} {Molecular Hydrodynamics}}}\
  (\bibinfo  {publisher} {Dover Publications, Inc.},\ \bibinfo {address} {New
  York},\ \bibinfo {year} {1991})\ \bibinfo {note} {reprint}\BibitemShut
  {NoStop}%
\bibitem [{\citenamefont {Straube}\ and\ \citenamefont
  {Höfling}(2023)}]{Straube:2023}%
  \BibitemOpen
  \bibfield  {author} {\bibinfo {author} {\bibfnamefont {A.~V.}\ \bibnamefont
  {Straube}}\ and\ \bibinfo {author} {\bibfnamefont {F.}~\bibnamefont
  {Höfling}},\ }\href@noop {} {\enquote {\bibinfo {title} {Depinning
  transition of self-propelled particles},}\ } (\bibinfo {year}
  {2023})\BibitemShut {NoStop}%
\end{thebibliography}%

\end{document}